\documentclass[a4paper,10pt,twoside]{cpc-hepnp}
\usepackage{multicol}
\usepackage{graphicx}
\usepackage{booktabs}
\usepackage{subfigure}
\usepackage{amssymb,bm,mathrsfs,amscd}
\usepackage[tbtags]{amsmath}
\usepackage{lineno}
\begin{document}

\fancyhead[c]{\small Chinese Physics C~~~Vol. xx, No. x (2021) xxxxxx}
\fancyfoot[C]{\small 010201-\thepage}

\footnotetext[0]{Received 8 Jan. 2021}

\title{Performance of LHAASO-WCDA and Observation of Crab Nebula as a Standard Candle }

\author{
F. Aharonian$^{26,27}$,
Q. An$^{4,5}$,
Axikegu$^{20}$,
L.X. Bai$^{21}$,
Y.X. Bai$^{1,3}$,
Y.W. Bao$^{15}$,
D. Bastieri$^{10}$,
X.J. Bi$^{1,2,3}$,
\\
Y.J. Bi$^{1,3}$,
H. Cai$^{23}$
% ...
\\
(The LHAASO Collaboration)
}
\maketitle

\address{%
$^1$ Key Laboratory of Particle Astrophyics \& Experimental Physics Division \& Computing Center, Institute of High Energy Physics, Chinese Academy of Sciences, 100049 Beijing, China\\
$^2$University of Chinese Academy of Sciences, 100049 Beijing, China\\
$^3$TIANFU Cosmic Ray Research Center, Chengdu, Sichuan,  China\\
$^4$State Key Laboratory of Particle Detection and Electronics, China\\
$^5$University of Science and Technology of China, 230026 Hefei, Anhui, China\\
$^6$Department of Engineering Physics, Tsinghua University, 100084 Beijing, China\\
$^7$National Astronomical Observatories, Chinese Academy of Sciences, 100101 Beijing, China\\
$^8$National Space Science Center, Chinese Academy of Sciences, 100190 Beijing, China\\
$^9$School of Physics, Peking University, 100871 Beijing, China\\
$^{10}$Center for Astrophysics, Guangzhou University, 510006 Guangzhou, Guangdong, China\\
$^{11}$School of Physics and Astronomy \& School of Physics (Guangzhou), Sun Yat-sen University, 519082 Zhuhai, Guangdong, China\\
$^{12}$School of Physical Science and Technology, Guangxi University, 530004 Nanning, Guangxi, China\\
$^{13}$Hebei Normal University, 050024 Shijiazhuang, Hebei, China\\
$^{14}$School of Physics and Microelectronics, Zhengzhou University, 450001 Zhengzhou, Henan, China\\
$^{15}$School of Astronomy and Space Science, Nanjing University, 210023 Nanjing, Jiangsu, China\\
$^{16}$Key Laboratory of Dark Matter and Space Astronomy, Purple Mountain Observatory, Chinese Academy of Sciences, 210023 Nanjing, Jiangsu, China\\
$^{17}$Institute of Frontier and Interdisciplinary Science, Shandong University, 266237 Qingdao, Shandong, China\\
$^{18}$Key Laboratory for Research in Galaxies and Cosmology, Shanghai Astronomical Observatory, Chinese Academy of Sciences, 200030 Shanghai, China\\
$^{19}$Tsung-Dao Lee Institute \& School of Physics and Astronomy, Shanghai Jiao Tong University, 200240 Shanghai, China\\
$^{20}$School of Physical Science and Technology \&  School of Information Science and Technology, Southwest Jiaotong University, 610031 Chengdu, Sichuan, China\\
$^{21}$College of Physics, Sichuan University, 610065 Chengdu, Sichuan, China\\
$^{22}$Key Laboratory of Cosmic Rays (Tibet University), Ministry of Education, 850000 Lhasa, Tibet, China\\
$^{23}$School of Physics and Technology, Wuhan University, 430072 Wuhan, Hubei, China\\
$^{24}$School of Physics and Astronomy, Yunnan University, 650091 Kunming, Yunnan, China\\
$^{25}$Yunnan Observatories, Chinese Academy of Sciences, 650216 Kunming, Yunnan, China\\
$^{26}$Dublin Institute for Advanced Studies, 31 Fitzwilliam Place, 2 Dublin, Ireland \\
$^{27}$Max-Planck-Institut for Nuclear Physics, P.O. Box 103980, 69029  Heidelberg, Germany \\
$^{28}$ Dipartimento di Fisica dell'Universit\`a di Napoli   ``Federico II'', Complesso Universitario di Monte
                  Sant'Angelo, via Cinthia, 80126 Napoli, Italy. \\
$^{29}$Institute for Nuclear Research of Russian Academy of Sciences, 117312 Moscow, Russia\\
$^{30}$Moscow Institute of Physics and Technology, 141700 Moscow, Russia\\
$^{31}$D\'epartement de Physique Nucl\'eaire et Corpusculaire, Facult\'e de Sciences, Universit\'e de Gen\`eve, 24 Quai Ernest Ansermet, 1211 Geneva, Switzerland\\
$^{32}$Department of Physics, Faculty of Science, Mahidol University, 10400 Bangkok, Thailand\\
}

\begin{abstract}
The first Water Cherenkov detector of the LHAASO experiment (WCDA-1) has been operating since April, 2019. The first 10
months of data have been analyzed to test its performance by observing the Crab Nebula as a standard candle.
The WCDA-1 achieves the sensitivity of 65 mCU per year with a statistical threshold of 5 $\sigma$. In order to do so, 97.7\% cosmic ray background rejection rate around 1 TeV and 99.8\% around 6 TeV with an approximately photon acceptance about 50\% by using the $compactness$ of the shower footprints to be greater than 10 as the discriminator between gamma induced showers and the cosmic ray backgrounds. The angular resolution is measured using the Crab Nebula as a point source about 0.45$^\circ$ at 1 TeV and better than 0.2$^\circ$ above 6 TeV with the
pointing accuracy better than 0.05$^\circ$. They are all matching the design specifications. The energy resolution is found 33\% for gamma rays around 6 TeV.
The spectral energy distribution of the Crab Nebula in the range from 500 GeV and 15.8 TeV is
measured and in agreement with results of other TeV gamma ray observatories.
\end{abstract}

\begin{keyword}
LHAASO-WCDA, Crab Nebula, angular resolution, spectral energy distribution
\end{keyword}

\begin{pacs}
PACS codes ( 95.85.Pw, 96.50.sd, 98.70.Sa)
\end{pacs}

\footnotetext[0]{\hspace*{-3mm}\raisebox{0.3ex}{We are grateful to all source of financial support for the corresponding research work. They are listed in the acknowledgement section.}}


\section{Introduction}

The Large High Altitude Air Shower Observatory (LHAASO)~\cite{cao2014} is a complex of
extensive air shower (EAS) detector arrays,  located at Mt.
Haizi (29$^\circ$21'27.6" N, 100$^\circ$08'19.6" E) and the altitude of 4410 m above sea level (a.s.l.),
in an area at the edge of the Qingzang plateau  near Daocheng, Sichuan Province, China.
It consists of an extensive air shower detector array covering an
area of 1.3 km$^2$ (KM2A) with 5195 scintillator counters (ED, 1 m$^2$ active area) and 1188 muon detectors (MD, water Cherenkov detector with an area of 36 m$^2$ buried under 2.5 meters of dirt).
In the center of the array are the Water Cherenkov Detector Array (WCDA) covering 78,000 m$^2$  and Wide Field-of-view air Cherenkov/fluorescence Telescope Array (WFCTA) of 18 telescopes.
LHAASO is designed for detection of air showers induced by all kinds of cosmic ray particles with energy ranging from a few tens of GeV to a few EeV, and for very effectively identifying gamma rays out of the
charged cosmic ray background. LHAASO is, therefore, a multi-purpose facility for very high energy gamma ray
astronomical observation, cosmic ray energy spectrum measurements for individual species and other wide-ranging topics associated with the cosmic rays or air shower phenomena.
For the gamma ray source survey, the designed sensitivity is about 1\% Crab Unit (CU).
Its wide FoV allows LHAASO to survey 1/7 of the northern sky at any moment for gamma ray sources.
With the operation in full duty cycle, LHAASO scans the entire northern hemisphere every 24 hours.
The spectra of all sources in its FoV will be measured with high precision over a wide energy
range from $10^{11}$ eV to $10^{15}$ eV.
These measurements, particularly above 100 TeV, will offer a great
opportunity for finding cosmic baryon PeVatrons and therefore identifying cosmic ray origins
among the gamma ray sources.
The Cherenkov/fluorescence telescopes of WFCTA are used to record the longitudinal development
of the air showers that trigger the arrays on the ground.
The combined detection of showers using all detectors in LHAASO enables the identification of
the cosmic ray species, at least for protons, H+He and iron nuclei with an aperture greater
than 4000~m$^2\cdot$sr. The knees of the spectra  of cosmic rays are expected to be measured at
energies above 100 TeV for individual mass groups.

The Water Cherenkov Detector Array (WCDA), mainly used for surveying transient  phenomena and
discovering new sources,
has been built in 3 phases. The first one, denoted as  WCDA-1, is a
square pond of 150 m$\times$150 m consisting of 900 detector units.
It has been completed in April 2019 and in operation since then. In the second phase, one more
pond of the same size, referred to as WCDA-2, has also been in operation since November 2019.
The third pond, WCDA-3, with a size of 300 m$\times$110 m, began operation at beginning of 2021. For gamma-ray induced showers,
WCDA-2 and WCDA-3 have a threshold below 100 GeV because they use a 20" PMT in each unit, while 600 GeV is the threshold for WCDA-1 because an 8" PMT is equipped in each unit. The threshold could be treated as such an energy above which the detector has a sensitivity $>1$ CU.

In this paper, the performance of WCDA-1 is thoroughly studied using the data taken from April 2019 to
March 2020. The experiment, including the detector configuration and the calibration
for both timing and charge measurements, is described in  the second section.
The details about the observation of Crab Nebula as a standard candle are described in the third section, including the shower geometric reconstruction, the cosmic ray background suppression and angular resolution. The details about air shower and detector response simulation and the gamma-ray shower energy reconstruction are
discussed in the fourth section.
The  measurement of the spectral energy distribution of the Crab Nebula is described along with the
systematic uncertainties in the fifth section. A short summary follows at the end of the paper.

\section{The WCDA-1 Detector Array}
\label{sec:experiment}

\begin{figure}
\centering\includegraphics[width=0.45\linewidth]{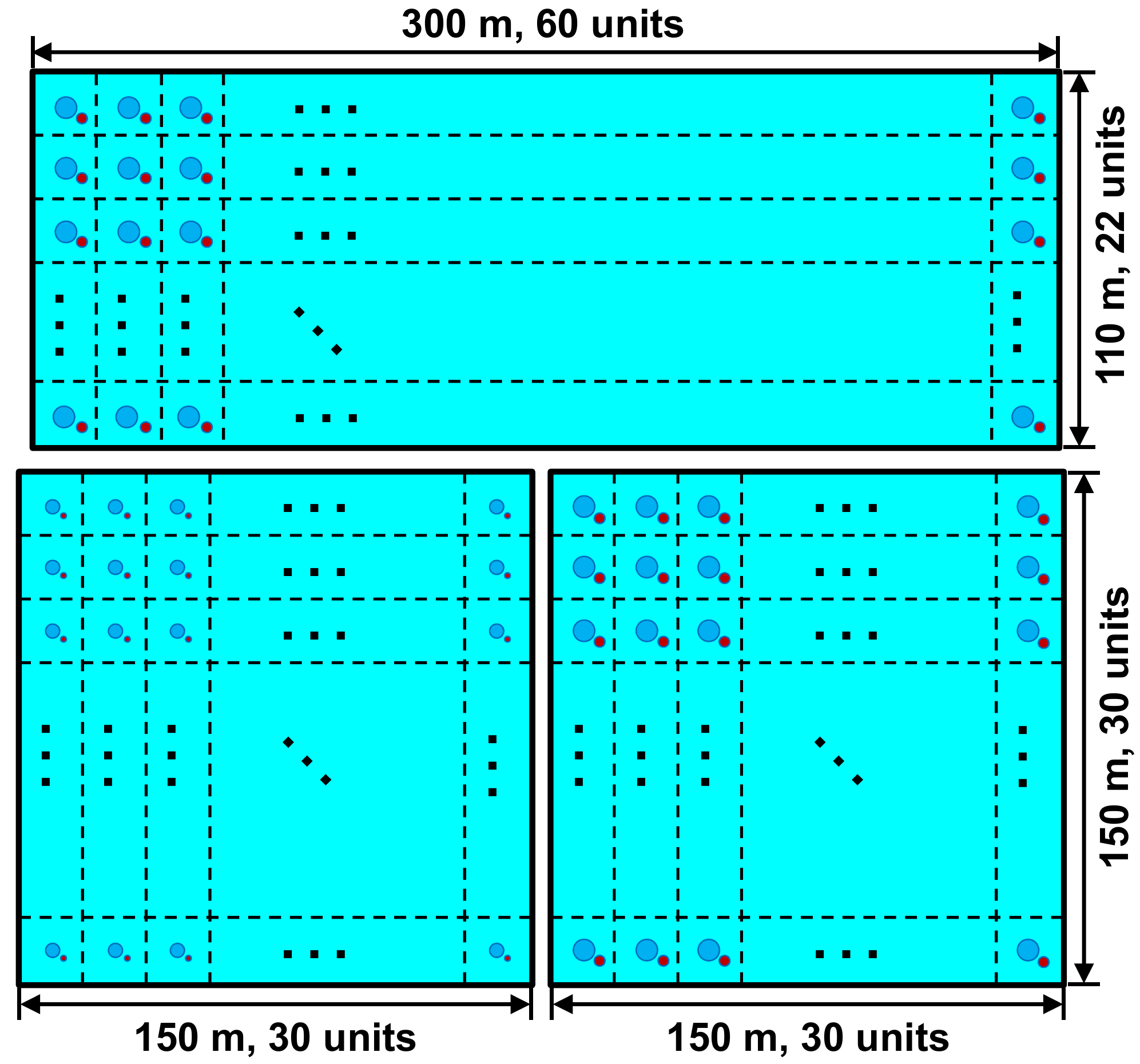}
\caption{Schematic of the WCDA layout. The lower two ponds are
WCDA-1 and WCDA-2 from left to right, respectively, and the upper one is WCDA-3. The two dots in each detector unit indicate the PMTs, while the dashed lines represent the curtains between units. Two combinations of PMTs are used:  8" and 1.5" PMTs in WCDA-1 and  20" and 3" PMTs in WCDA-2 and WCDA-3.}
\label{wcdalayout}
\end{figure}

WCDA-1 is composed of 900 detector units optically separated by non-reflecting black plastic curtains.
Each unit is a 5 m $\times$ 5 m cell equipped with two upward-facing PMTs on the bottom at the center of
the unit. The pond is filled with purified water up to 4 m above the photo-cathodes
of the PMTs, with a total volume of about 100,000 cubic meters.
This is the first completed detector array out of three, as illustrated in the layout
in Fig.~\ref{wcdalayout}. The inner surface of the pond is covered with high density
polyethylene (HDPE) film with a thickness of 2~mm
in order to guarantee a daily leakage rate less than 0.02\% of the total water capacity.

To maintain an attenuation length for near-ultra violet light
longer than 15 meters, the water needs to be purified continuously. This is done with a closed recycling loop.  Water is sucked out from one side and filled back in on the opposite side of the pond after purification with bacterium/organic-carbon killing and filtering. The whole volume is recycled every 15 days.

\begin{figure}[t]
\centering
{\includegraphics[width=12cm]{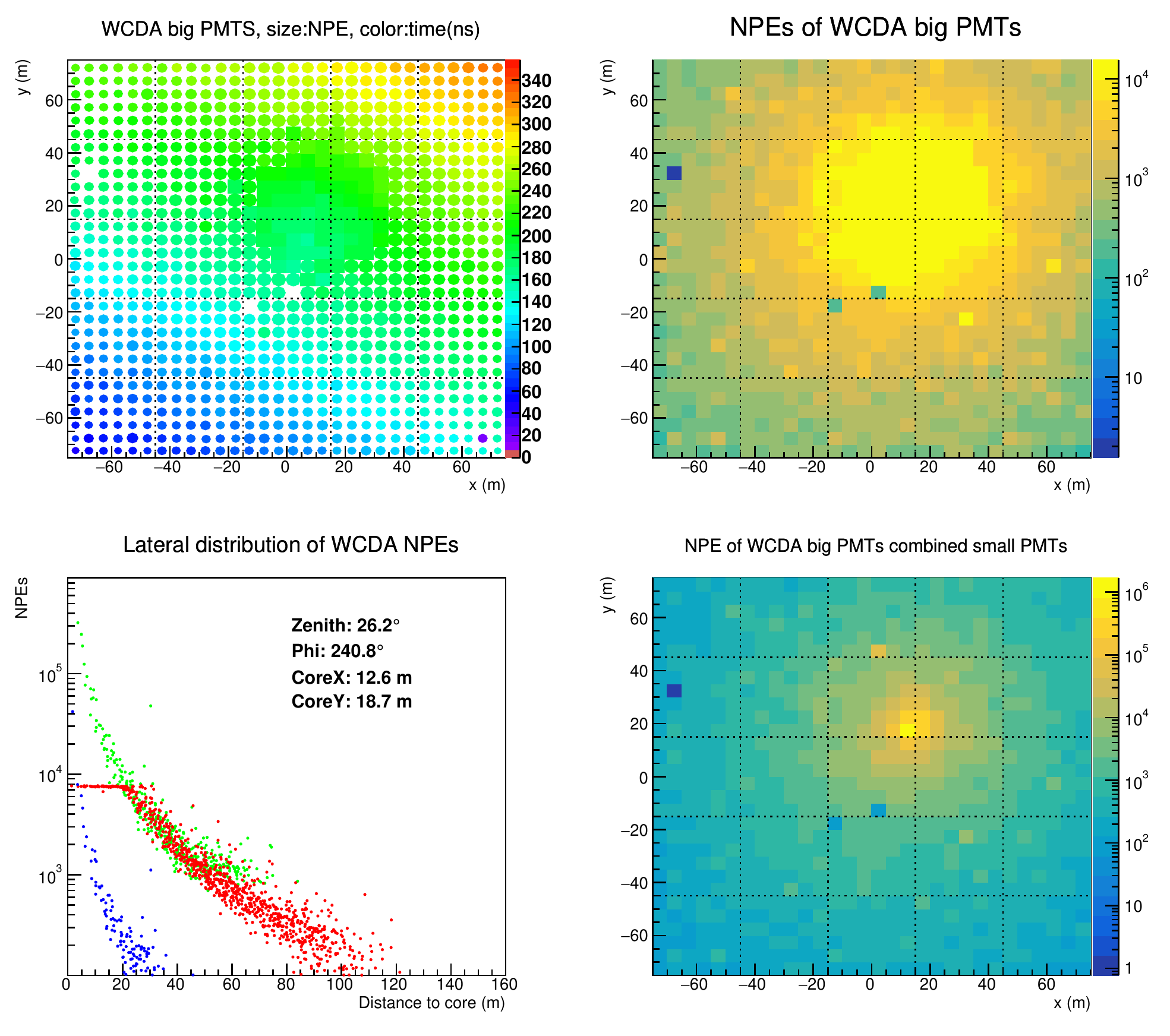}}
\caption{A typical high energy shower event detected by WCDA-1. The top left panel shows the arrival time of the shower front at each cell. The units on the colour scale are ns. The top right panel shows the number of photoelectrons, N$_{PE}$, recorded by
the 8" PMT in each cell (bin) in which many saturated PMTs can be seen.
To improve the measurement of the core region, the two maps are combined, as shown in lower left panel,
where the lateral distribution, measured in equivalent N$_{PE}$ , of  the small PMTs (blue points represent its original measurement and green points represent the charge equivalent to the charge that would be measured by the  8" PMT) is
overlaid to that of the big PMTs (red points) to construct a distribution over the whole dynamic range.
The lower right panel shows the same event, where the information of the 2 PMTs is combined. }
\label{event}
\end{figure}

In Fig.~\ref{wcdalayout},  two dots in each unit indicate two PMTs of 8" and 1.5", respectively, used in WCDA-1.
The combination of the two covers a wide dynamic range in number of photoelectrons (PE) from 1 to 200,000.
This configuration allows a better measurement of the particle density distribution in the shower
cores without saturation even for very energetic showers up to 10~PeV.
It also allows the shower core location to be reconstructed with a resolution better than 3~m for 10 TeV and more energetic events.  This  greatly improves the shower energy
resolution for those events. It is useful for high precision measurements of spectral energy distribution
(SED) of gamma-ray sources in the overlapping region with LHAASO-KM2A, and It is essential for measuring the charged cosmic ray spectrum in the combined experiment with LHAASO-WFCTA.
An event detected by both large and small PMTs is shown in Fig.~\ref{event}.
It clearly shows how the combination of the charge measurements (in units of PE) from the two PMTs
correctly measures the lateral distribution for high energy showers.

The 8" PMTs are read out from both the anode and the last dynode as two independent signals with a
gain difference of a factor of about 48.
Two 12-bit ADC channels are used to digitize both signals and achieve a dynamic range from 1 PE
to 5000 PE with a non-linearity less than 1.5\% and a charge resolution of 35\% for
single PE and 2\% for signals greater than 200 PE.
The anode signal is amplified and fed into a discriminator to measure the trigger time.
A timing resolution of 1 ns has been achieved, enough to reconstruct the shower front conical structure.
The 900 units in WCDA-1 are divided into 25 groups of 36 units (72 PMTs),  each  connected to a single electronic box, which provides the power and hosts  all electronics boards.

The White Rabbit Protocol (WR)\cite{WR-LHAASO} is
used to distribute the clock and guarantee the synchronization of each detector unit within 200 ps through a fibre network over the  entire LHAASO array.
In the central timing station, a rubidium oscillator synchronized with GPS/BeiDou Time (BDT) is the source of the clock, which is  distributed to  the whole array through the WR network.
Data collected from each PMT are transmitted to the data station of WCDA-1,  through the same fiber network. The same is true for any other LHAASO detector, where the same network is used for synchronization and data readout.
In WCDA-1, the correct synchronization of the 36 large PMTs,  used for timing measurements within each cluster,
is achieved with a dedicated calibration described below.

A hit is formed with the threshold of 1/3 PE  for a 8" PMT and
recorded with the absolute time stamp from the WR system and charges simultaneously measured by the 8" PMT and the 1.5" PMT.
The stream of hits is transmitted to the data acquisition (DAQ) computers and buffered for 30 seconds before being discarded.
The hits are sorted in time and fed into the triggering and event building algorithm.
This  maximises the flexibility by allowing many trigger algorithms to run in parallel.
Built events are stored for further analysis in specific data streams according to the trigger algorithms.

\subsection{Timing and Charge Calibration}

\begin{figure}
\centering
\subfigure[]{\includegraphics[width=0.6\linewidth]{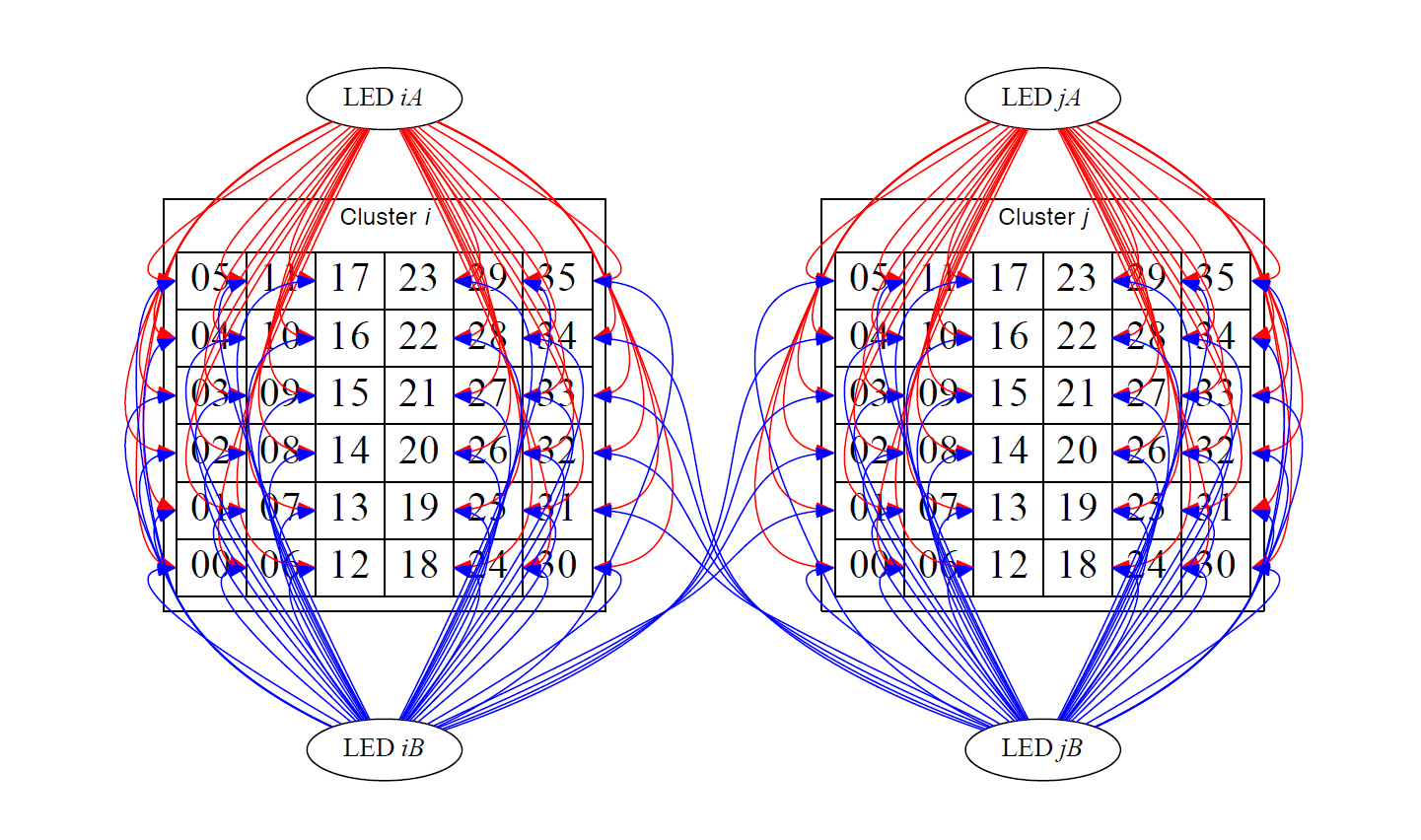}}\\
\subfigure[]{\includegraphics[width=0.4\linewidth]{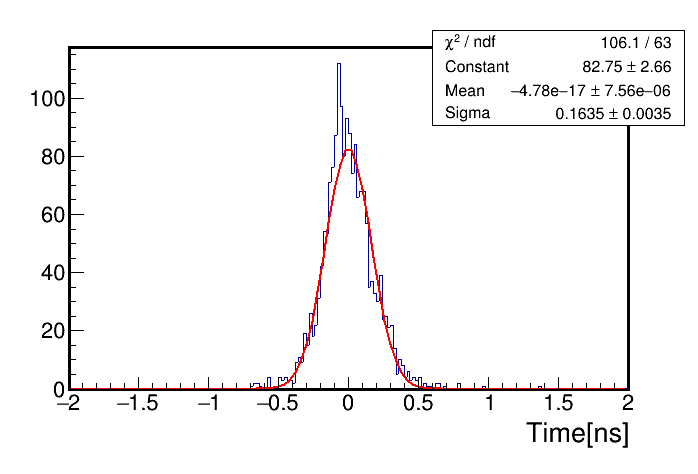}}
\subfigure[]{\includegraphics[width=0.4\linewidth]{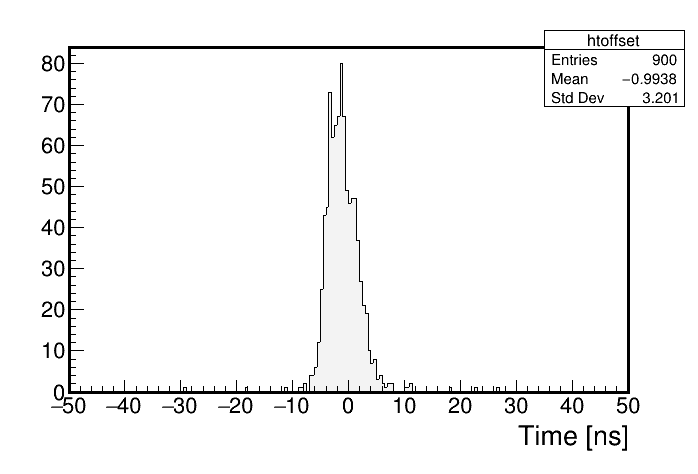}}
\caption{(a) A sketch map of time cross calibration.
(b) The distribution of the time offset for 1800 fibers used in the
calibration system of WCDA-1.
(c) The distribution of the time spread among the 900 units of WCDA-1.}
\label{timecalib1}
\end{figure}

In the WCDA-1 architecture, 900 units are divided into 25 groups. The 36 8" PMTs in each group are connected to an electronic box, which is synchronized with the WR system.  In order to calibrate the 36 PMTs in a group, an  optical fiber bundle is illuminated by a LED and used to distribute synchronized light pulses to every PMT in the group.
The LED is driven by a pulse generator installed in the electronics box.
Before routing the fibers to PMTs, the arrival time of the light pulse
generated by the LED through each fibre is measured~\cite{bundle-calib-LJY}  in a lab with an
accuracy of better than 100 ps. The time differences between fibers are shown in  Fig.~\ref{timecalib1}  (b) as a distribution with a Root Mean Square (RMS) of 163 ps among 1800 fibers.
In order to check the synchronization between the different groups,
another bundle of 36 fibers is also installed in each group. Twenty fibers of this bundle
are used to illuminate 20 PMTs belonging to the group, and the remaining 16 fibers
illuminate 4 PMTs in each of the 4 adjacent groups as shown in Fig.~\ref{timecalib1} (a).
The Root Mean Squared (RMS) time offsets measured between adjacent groups is 200 ps. This confirms
the synchronization accuracy achieved by using the WR system.

The calibration using the fiber network has been carried out on site several times, including the first one during
the `dry-run' before water filling and those in March, April, May,
August and October 2019.
The distribution of time offsets  measured in the last calibration run is shown in  Fig.~\ref{timecalib1} (c). The RMS of the offsets is 3.2 ns among
the 900 PMTs. The calibration data and the calibration system have proven to be very
stable over time.  The time offsets are very stable with a maximal variation less than 0.2 ns among 900 PMTs over all the calibrations over eight months.

\begin{figure}
\centering
\subfigure[]{\includegraphics[width=6cm]{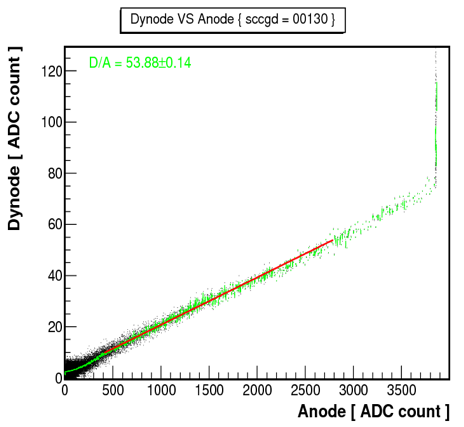}}
\subfigure[]{\includegraphics[width=6.5cm]{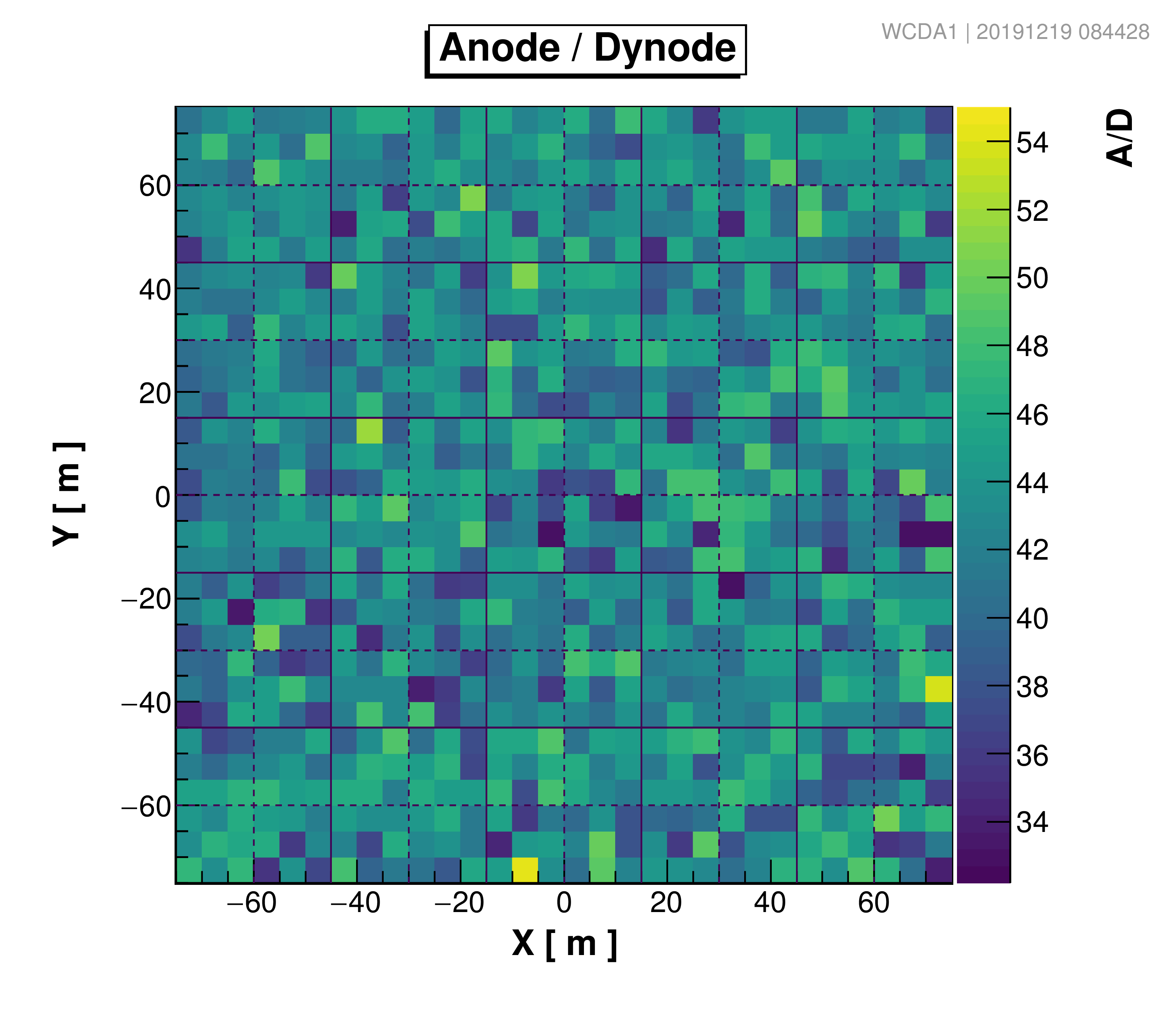}}
\caption{(a):  An example of the amplitude, in ADC counts, of the last dynode  versus the anode signal for an 8" PMT.
(b): Uniformity over the WCDA-1, as a 2D map,  of the measured A/D ratios for all 900 PMTs.}
\label{charge1}
\end{figure}

As mentioned above, the 8" PMT is read out from both the anode and the last dynode to cover a range from 1/3 PE to 5000 PE.  The charge calibration of the anode signal is straightforward by measuring the  single PE peak position of the ADC count distribution for each PMT. The resolution of the single PE position, with 25 ADC counts as the typical value, is about 32\%. The range covered by the anode high gain channel is up to 130 PE with non-linearity less than 5\%. The details of the calibration are reported elsewhere \cite{PMT-test} in which the measurements of the linearity of
every PMT have been summarized.
The calibration of the dynode signals is
done by comparing them with the corresponding
anode signals in the overlapping region, namely above 25 PE.
In Fig.\ref{charge1} (a), the dynode signal is plotted versus the anode signal
in terms of ADC counts for a typical PMT. There exists saturation when the anode signal
exceeds 4096. The ratio between the two signals, denoted as $A/D$, could be
extracted by fitting in the region of the anode signals between 500 and  2300 counts.
The  ratio found, $A/D=53.9\pm 0.1$, is used as a calibration constant for the PMT.
A 2-D map of the 900 $A/D$ ratios of WCDA-1 is shown in the right panel of
Fig.~\ref{charge1}.
To investigate the calibration stability, the single PE peak position
and $A/D$ of all 900 PMTs  have been monitored continuously. The temporal variations were found
to be within 0.1\% and 0.05\%, respectively, during April and May 2019.

\begin{figure}
\centering
\subfigure[]{\includegraphics[width=7.2cm]{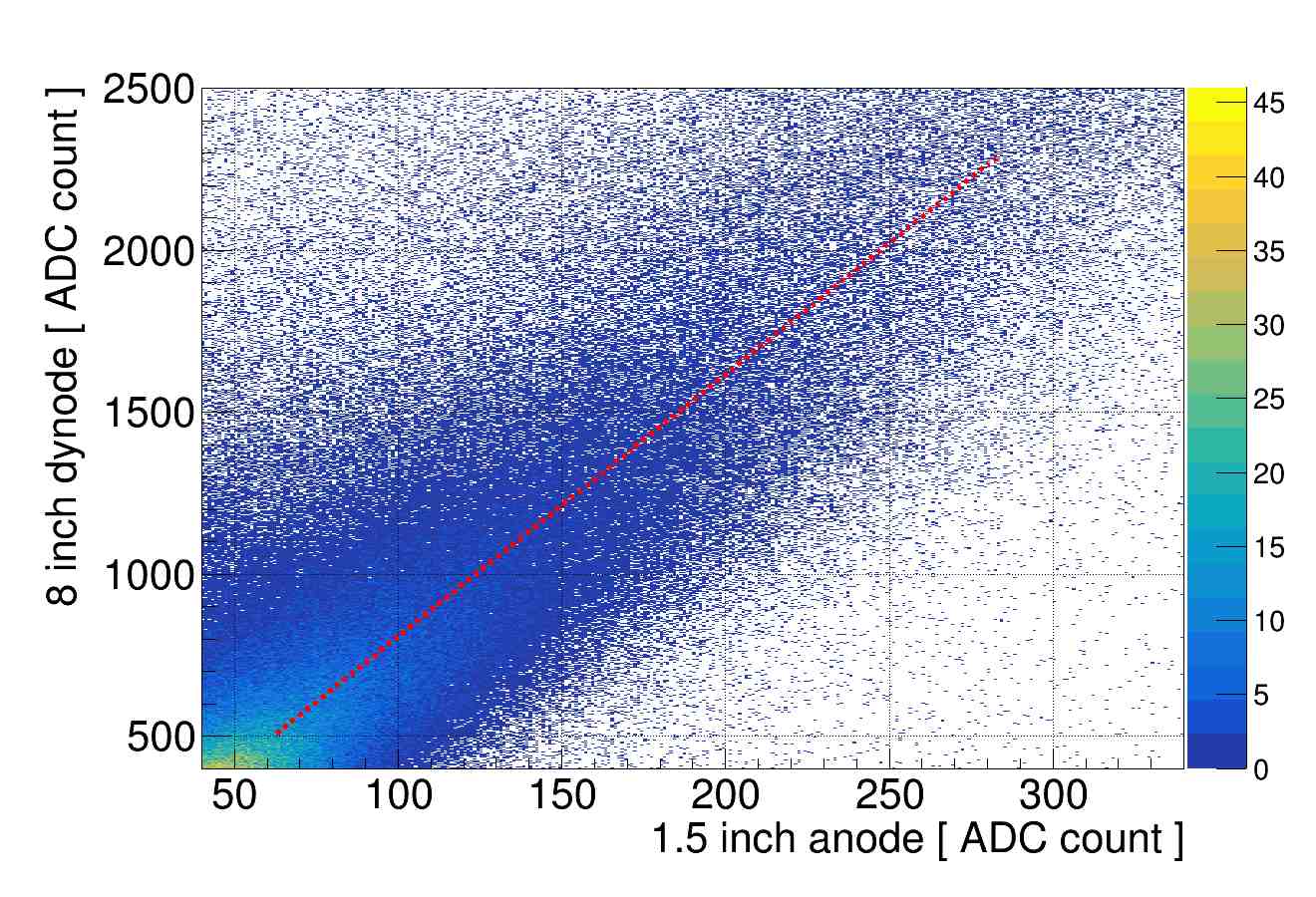}}
\subfigure[]{\includegraphics[width=7cm]{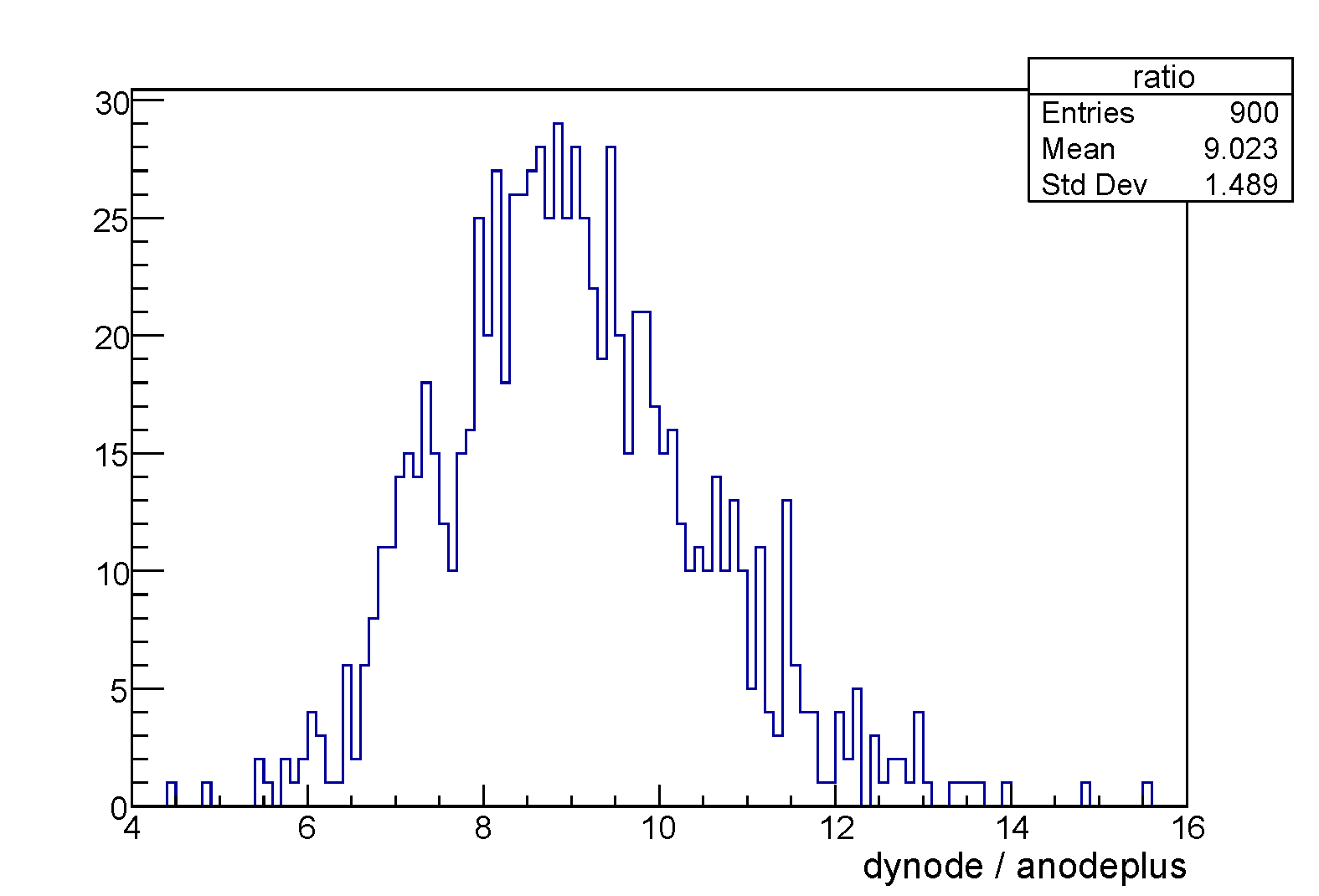}}
\caption{(a): Amplitude in ADC counts of dynode signal of 8" PMT versus anode signal
  of 1.5" PMT in a typical PMT unit of the WCDA-1 detector. (b): This measured A/D ratio
for all the 900 PMTs.}
\label{charge2}
\end{figure}

The calibration of 1.5" PMTs has been done by matching the signals
with the 8" PMTs in the overlapping region, i.e., above 500 PE measured by the 8" PMTs.
Fig.~\ref{charge2}(a) shows an illustrative example of
how the ratio of the two  signals  is extracted with a linear fit in the overlapping region,
between 500 and 2300 ADC counts of the  8" PMT, where the non-linearity is below 3\%.
The ratio is $8.18\pm0.04$ for the pair of PMTs as a example.
The distribution of the ratios of all 900 pairs has a width
of 17\%.
The mean value of 9.0 is related the ratio of the geometrical areas of the photocathodes of the two PMTs and their gains.

\begin{figure}
\centering
\includegraphics[width=8cm]{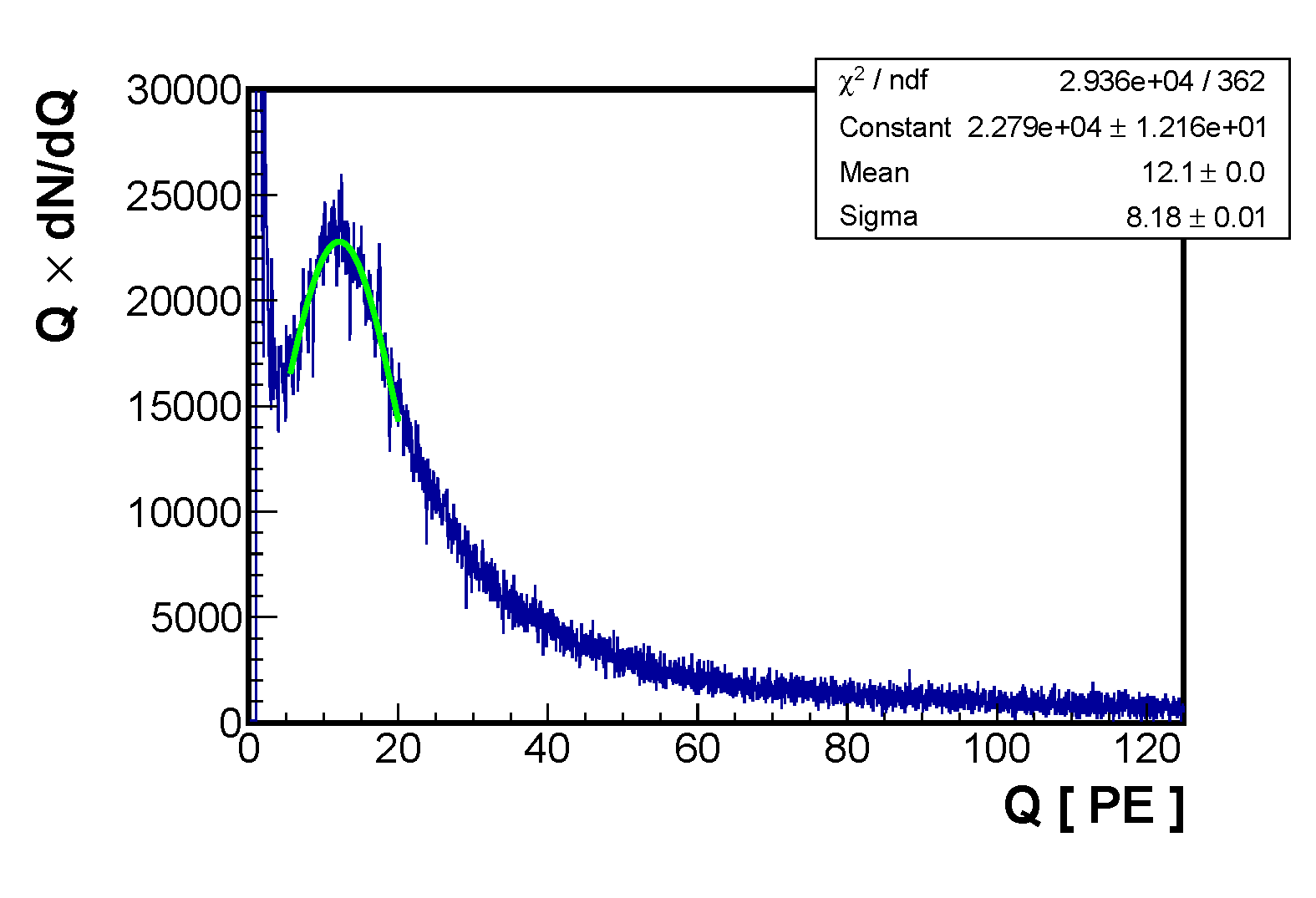}
\caption{The ``spectral charge distribution" of cosmic rays measured by a typical WCDA-1 unit. The clear ``muon peak" is due to the diffuse single muons falling in the unit.}
\label{muon_peak}
\end{figure}

\begin{figure}
\centering
\subfigure[]{\includegraphics[width=6cm]{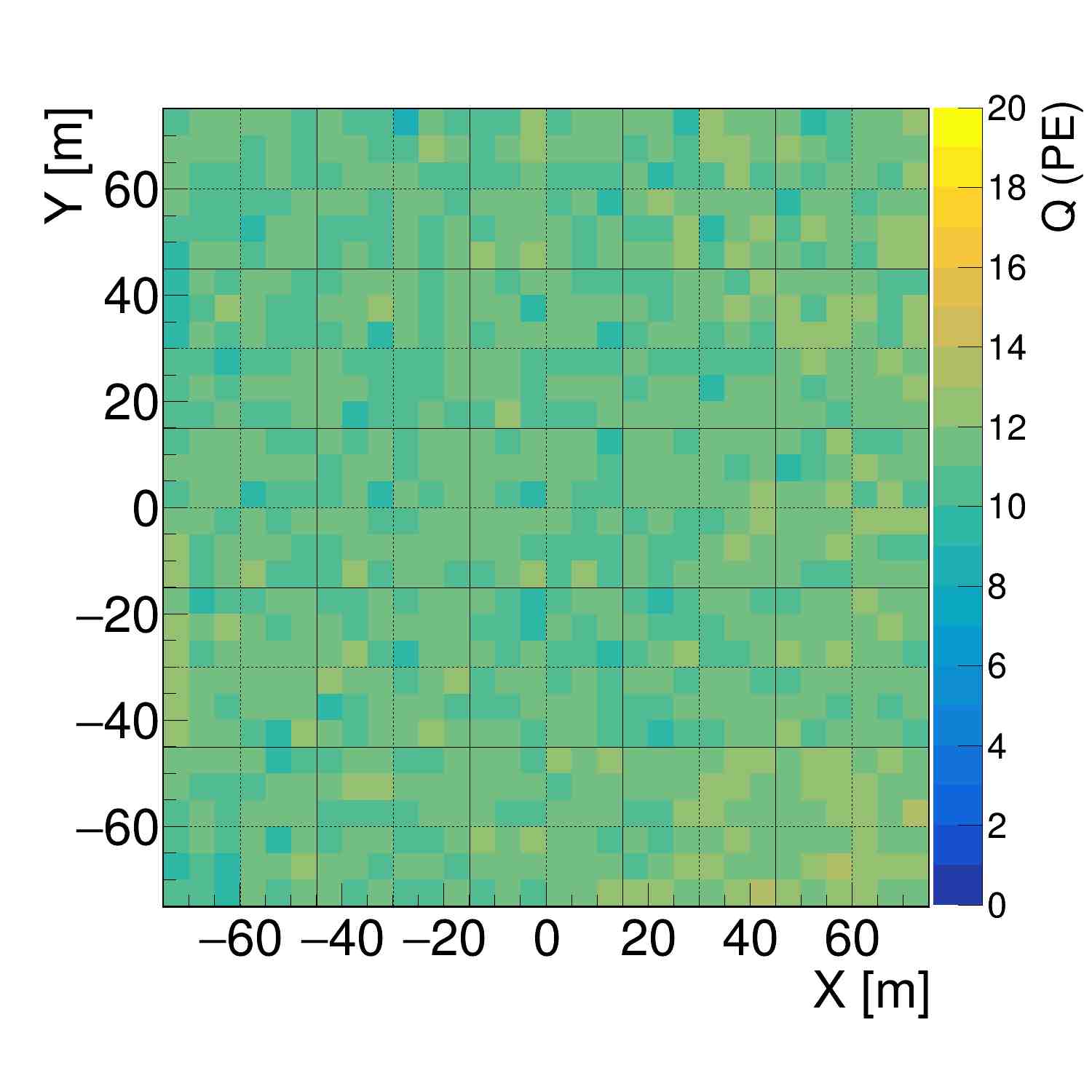}}
\subfigure[]{\includegraphics[width=8cm]{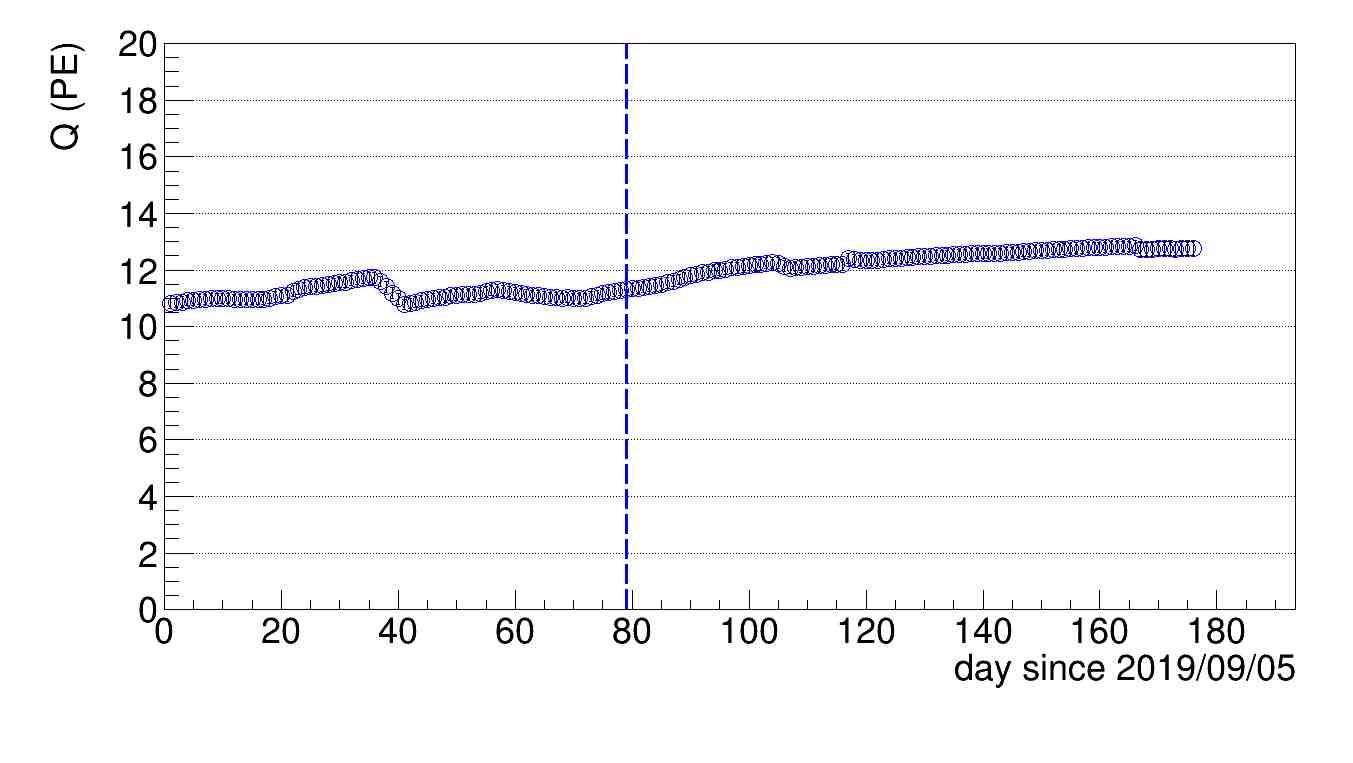}}
\caption{(a): Map of charges at the ``muon peak" measured by all WCDA-1 units on Nov. 22, 2019. The average charge is 11.6 PE and the spread of charges is 0.6 PE among the 900 units. (b): The variation of the average charge at the muon peak over 179 days from Sept. 5, 2019 to Feb. 29, 2020. This variation indicates the change of water transmission over this period of time. The dashed vertical line marks the  `standard' day Nov. 22, 2019.}
\label{muon-calib}
\end{figure}

The `spectral charge distribution' of cosmic rays measured by each WCDA-1 unit, i.e., $Q {\rm d}N/{\rm d}Q$ where $Q$ is the charge measured by the PMT in units of PE and $N$ is the trigger frequency of the unit by cosmic rays, is found to have a clear peak as shown in Fig.\ref{muon_peak} for a typical unit. This is due to the diffuse single cosmic ray muons which arrive at each unit with a distribution that is uniform and nearly constant from a wide range of directions. Studies show that the peak can be mainly attributed to muons passing by the PMT with a distance of about 3 meters~\cite{lihuicai}.  The measured charge at the peak indicates a combination of the PMT sensitivity and the transmission of water in the unit. This provides an ideal calibrating beam for WCDA-1 units to measure the uniformity in response of the 900 units and monitor the variation of the calibration with time, e.g., due to changes in water transmission. In the left panel of Fig.\ref{muon-calib}, the distribution of the charge at peak of 900 units is shown in the 2-D map of WCDA-1. The average charge of 900 peaks was 11.6 PE on Nov. 22, 2019. A distribution with the RMS of 0.6 PE measures the non-uniformity between the units in WCDA-1. The monitoring of the average charge of the peaks shows that the water transmission was continuously changing over 179 days since Sept. 5, 2019 when WCDA-1 began the stable operation phase.   In the right panel of Fig.\ref{muon-calib}, the average of the charge at peaks has been plotted as a function of time. All the differences between units and time variations have been corrected with respect to a ``standard detector" at Nov. 22, 2019 to maintain the uniformity and stability of the detector.

With such a well-calibrated uniform and stable detector, the trigger rates of detector units due to the high energy cosmic rays, instead of the muons, should be the same among the 900 units and also stable. This has been tested at the threshold $Q_{th}=2$ PE. It is found that the spread over 900 units is within 2.3\%. The stability of the rate is 1.3\% over 170 days since Sept. 1, 2019. This test also shows a spread of 5.9\% and stability of 2.9\% at $Q_{th}=0.3$ PE. This will be discussed in the systematic analysis below.

\section{Observation of Gamma Rays from the Crab Nebula for WCDA-1 Performance Testing }

\begin{figure}
\centering
\includegraphics[width=10cm]{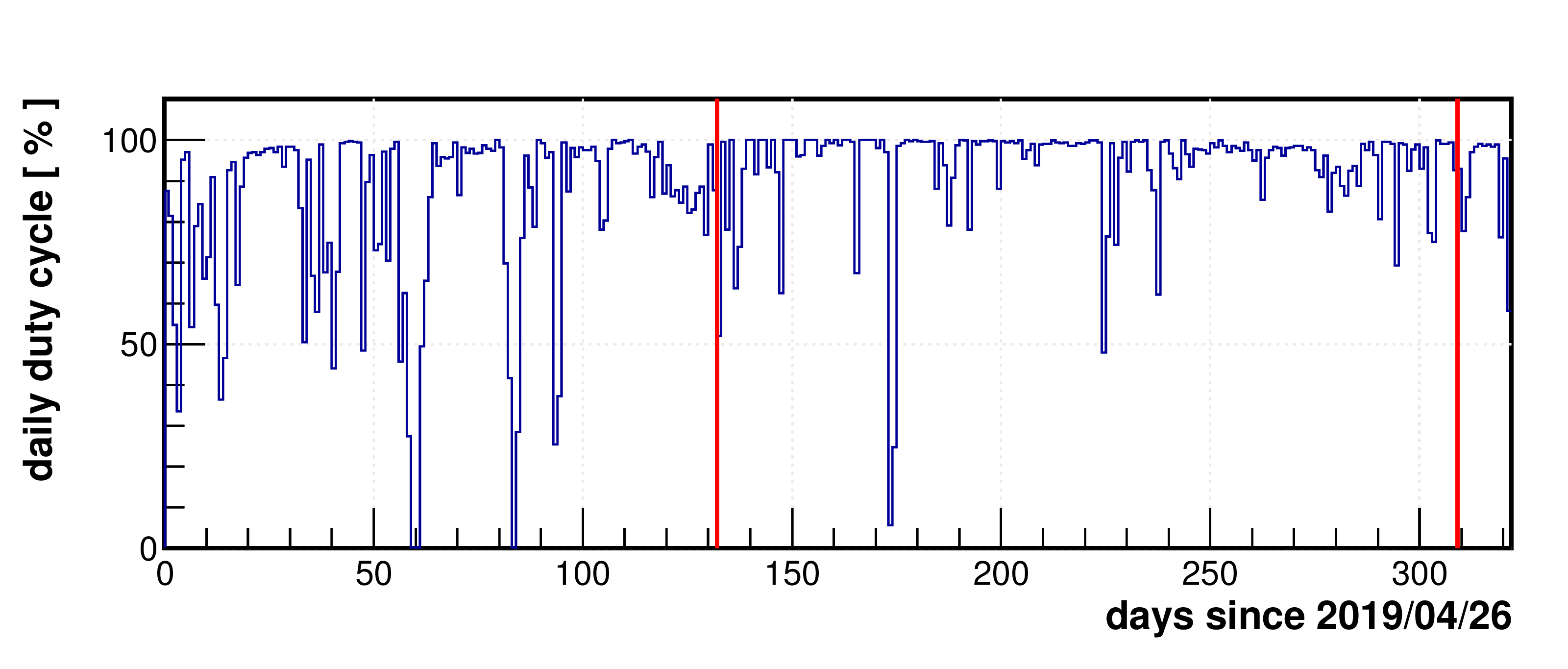}\\
\caption{The daily live time of WCDA-1  as a function of days since April 26, 2019. Two red vertical lines indicate the stable operation period, i.e. from Sept. 5, 2019 to Feb. 29, 2020, in which data are used for the analysis in this paper.   About 1.7$\times 10^9$ shower events are recorded and successfully reconstructed everyday  during this period of time. }
\label{tlive}
\end{figure}

WCDA-1 started operation on April 26, 2019. In a short test run at the beginning,  a large volume of monitoring data was collected to validate the
operation.
A trigger algorithm was implemented to record cosmic ray air showers by requiring at least 20 groups, a group being 3$\times$3 detector units, with each group being hit simultaneously in a window of 300 ns. A hit is formed if any one of the nine 8" PMTs in the cluster received a signal greater than the threshold of 1/3 PE.
The detector has been fully operational for stable survey of the northern sky since then.
The air shower event rate is about 45 kHz, while the trigger rate of a single detector unit is 22.5 kHz on average.
Once a trigger is formed, charges recorded by both 8" and 1.5" PMTs in 2 $\mu$s
are read out together with the absolute trigger time to build a complete shower event. Further reconstruction of the air shower for its geometrical properties is carried out after the event building. In Fig.\ref{tlive}, the daily live time of WCDA-1 is plotted as a function of time since April, 26, 2019, the day WCDA-1 was turned on.  $\thicksim 1.7\times10^{9}$ shower events are recorded and successfully reconstructed everyday during the stable operation for 179 days from Sept. 5, 2019 to Feb. 29, 2020. The analysis reported in this paper uses the data taken in this period of time. Afterwards, the operation was switched to the combined mode with both WCDA-1 and WCDA-2 participated.

\subsection{Reconstruction of Shower Geometry }
\label{geo-reconstruction}

For an air shower event, the arrival direction and core location are reconstructed in two steps.

The first step is to find the arrival direction by assuming a planar shower front using $N_{hit}$  units that are registered in the event with their locations $(X_i$,$Y_i)$ and trigger time $t_i$. The charge $Q_i$ is used to weight the $i$-th unit in a defined way. The arrival direction is the normal vector of the plane.
Units that have a trigger time too far ($>100$ ns) from the plane are considered to be random triggers by noise and rejected in further steps with the remaining  $N_{hit}$  units. The fitting procedure is iterated until the number $N_{hit}$ of firing PMTs becomes constant.

The second step is to find the shower core location $(X_c, Y_c)$ and refine the arrival direction reconstruction by assuming a conical
shower front centered at $(X_c, Y_c)$.  The core location is found by calculating the charge-weighted
averages \mbox{$X_c=\Sigma^{N_{10}}_{i=1} Q_iX_i/\Sigma^{N_{10}}_{i=1} Q_i$} and \mbox{$Y_c=\Sigma^{N_{10}}_{i=1} Q_iY_i/\Sigma^{N_{10}}_{i=1} Q_i$}, where $N_{10}=0.1N_{hit}$ and $i$ runs over the units that record the highest charges.
The conical front is defined by the time offsets $t(R)=c\cdot R$ with respect to the plane found in the previous step, where $R$ is the distance from the core in the plane and $c$ is the conical coefficient, set to be 0.07 ns/m for gamma ray induced showers in the working energy range of WCDA-1.
At this point, units that have trigger time farther than 30 ns from the conical front are rejected again as
noise triggers. $N_{hit}$ is renewed by counting the surviving hits.

\begin{figure}
\centering
\includegraphics[width=10cm]{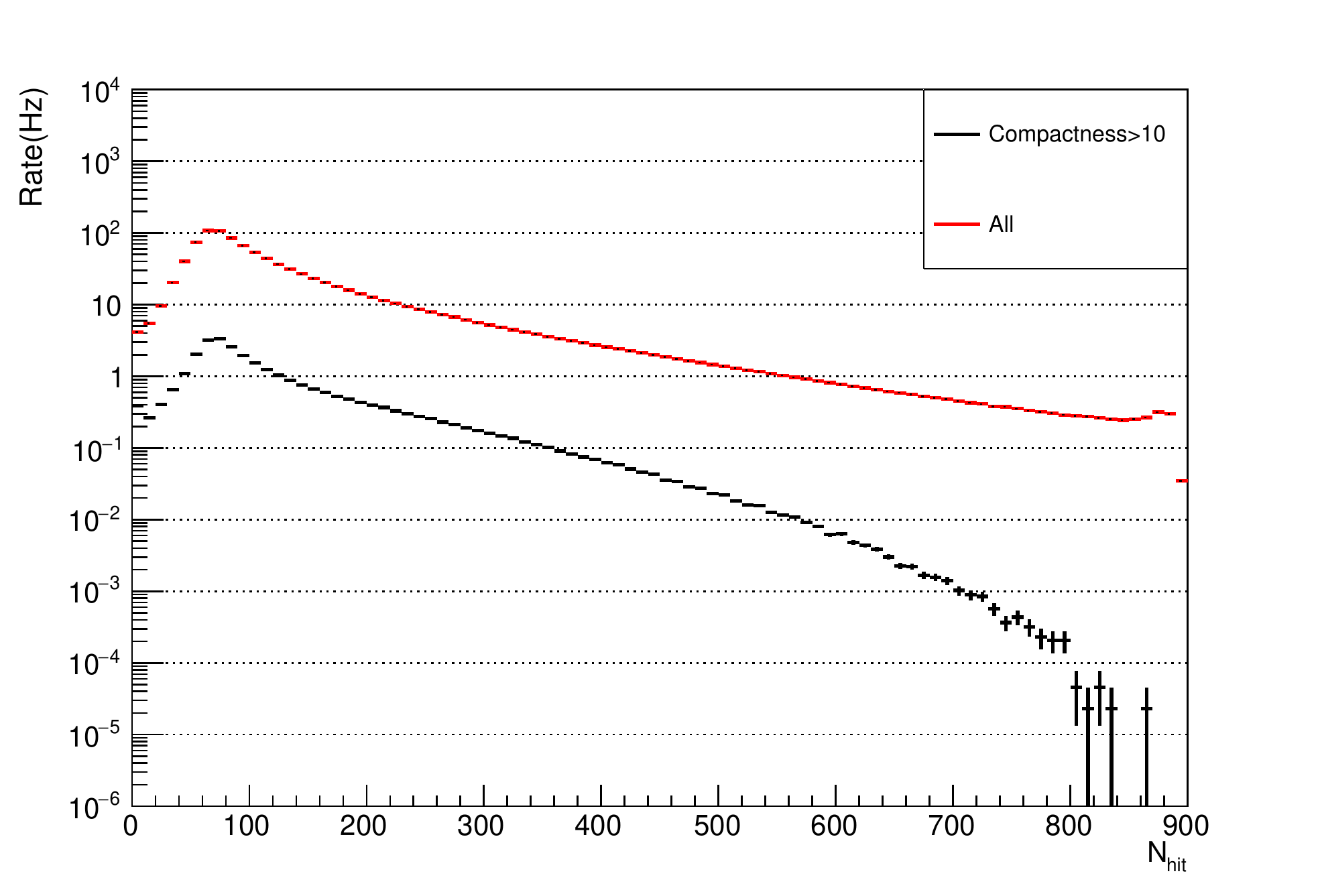}
\caption{Distribution of the
number of triggered detector units $N_{hit}$ after filtering out random noise hits (in red) and after the $C>10$ gamma selection cut (in black). This cut causes a reduction of 97.2\% for events with $N_{hit}<$ 550.}
\label{nhit-distr}
\end{figure}

$N_{hit}$ is selected as a shower energy estimator in the analysis. For showers in the WCDA-1 working energy range, most
of the hits have only one or two PEs. This enables a simple shower energy reconstruction based on the nearly linear relationship between $N_{hit}$  and the primary energy.
For the pure electromagnetic cascade induced by gamma rays, the Monte-Carlo simulation of the shower development is reliable to establish the energy response function.  The distribution of $N_{hit}$ of the well reconstructed shower events is shown in Fig.\ref{nhit-distr} as the curve in red. There is a clear pile-up of events with $N_{hit}>840$ that indicates the saturation of $N_{hit}$, which has a maximum of 900. This will
be discussed in detail in Sec. \ref{sec:E-recons} \rm.

\subsection{Cosmic Ray Background Suppression}

To maximize the sensitivity to gamma ray showers, an effective cosmic ray background
suppression is essential.

\begin{figure}
\centering
\includegraphics[width=6cm]{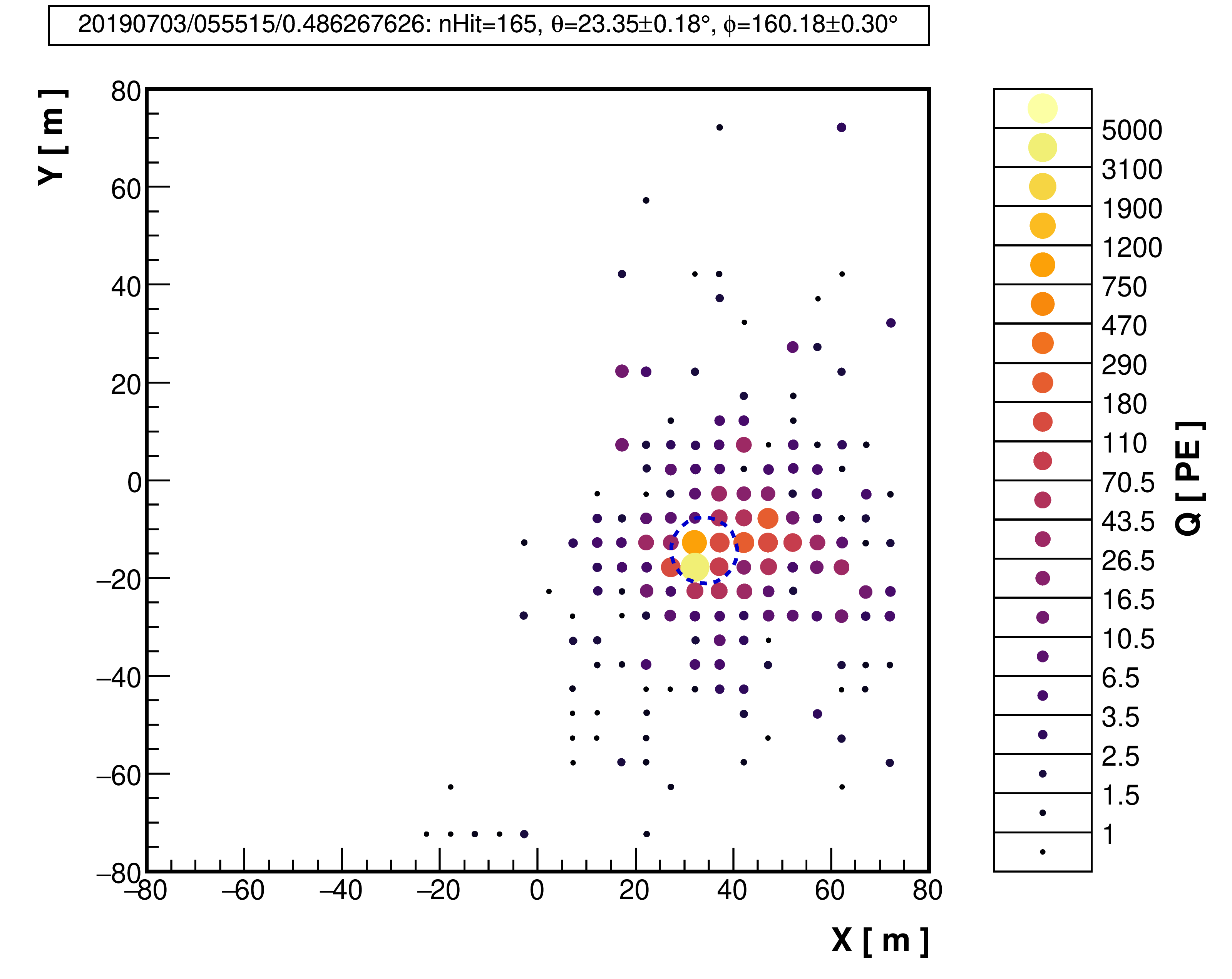}
\includegraphics[width=6cm]{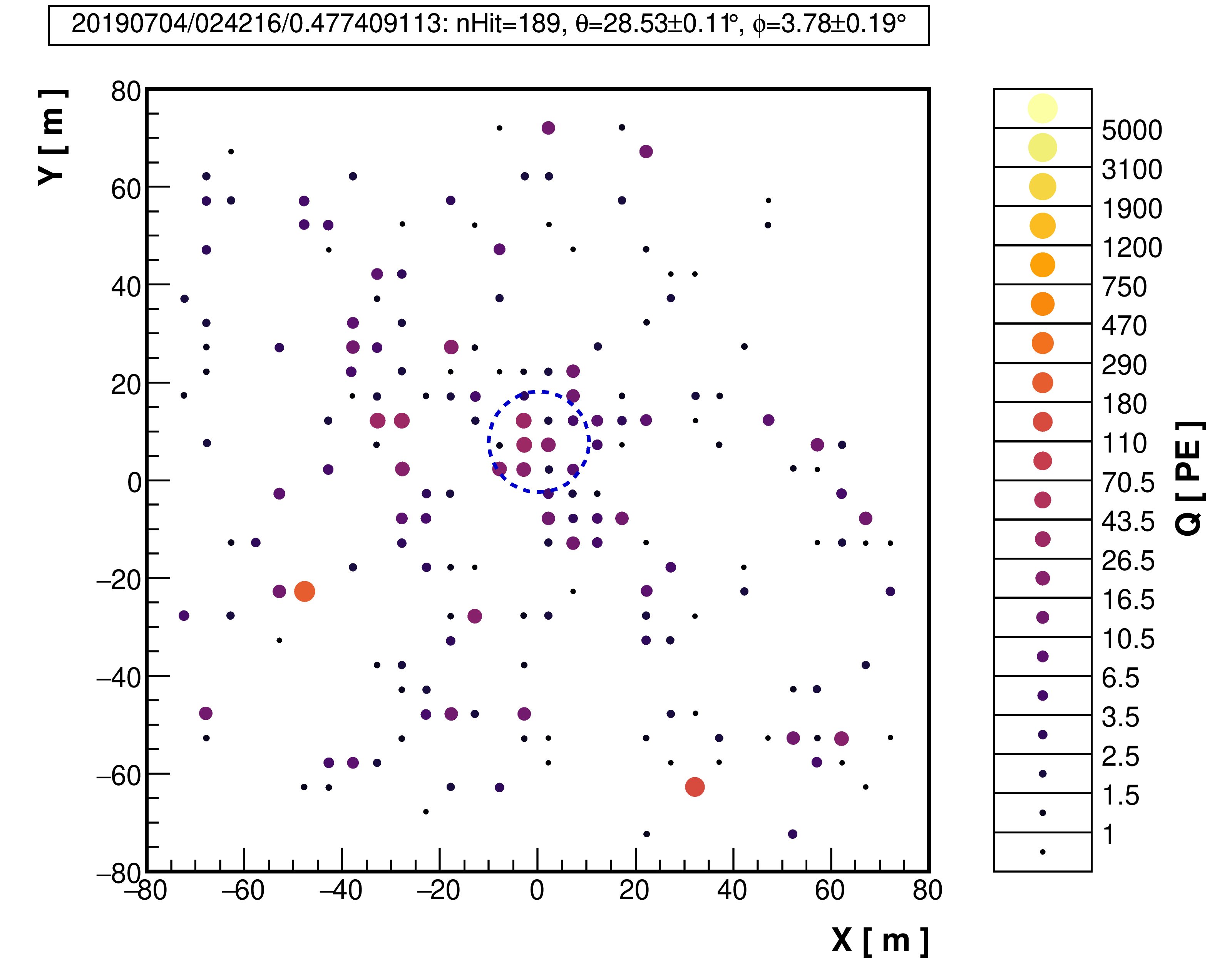}
\caption{A high-confidence gamma-ray candidate event from the direction of the Crab Nebula
  with $N_{hit}$=165 and a compactness value of $C=39$ (left panel)
  and a similar background cosmic-ray event with $N_{hit}$=189 and $C=1.0$ (right panel). Dashed circles indicate the reconstruction uncertainties of the shower cores.}
\label{gamma_like}
\end{figure}

For showers induced by cosmic rays, mainly protons, the secondary particles are more
spread out laterally than for gamma-ray-induced showers. In either case, secondary particles, mainly gamma rays and electrons, rapidly induce sub-showers in water and lose all their energy over a few tens of centimetres.
All relativistic charged particles in the sub-showers generate Cherenkov light in water
with the emitting angle of 41.3$^\circ$. The Cherenkov photons in each detector unit are sampled by the PMTs, thus
forming a footprint of the shower in WCDA-1. A typical cosmic ray event with $N_{hit}=189$ is shown in the right panel of Fig.\ref{gamma_like}. Each dot indicates a registered unit with the size indicating the corresponding measured charge. A clear dense area indicated by the dashed circle marks the core location of the shower; however, many large hits are also found throughout the footprint. Those `hot spots' are mainly due to
muons produced in the air shower which can easily penetrate through the entire depth of WCDA-1,
producing  Cherenkov light along the whole tracks, thus yielding stronger signals. Since muons are produced by the decay of $\pi^{\pm}$ generated in the shower, they carry the transverse momentum of pions so that they are distributed over a wider area in
general, when they arrive to the ground.
Therefore, on top of the smooth lateral distribution yielded by the electromagnetic component, there are isolated `hot spots' outside the core region.

In contrast, a typical gamma-ray-induced shower from the direction of the Crab Nebula as shown in the left panel
of Fig.\ref{gamma_like} has a more compact and rather smooth distribution of charges in the footprint measured by WCDA-1. With the similar  $N_{hit}=165$, the event has a better defined single core region because the shower is a  pure electromagnetic cascade process. Such  clear difference between the gamma ray signal and the cosmic ray background allows a way to suppress the flux of cosmic rays in a specific gamma ray source region, in particular for a large detector such as WCDA-1 that contains almost all of the secondary
particles in showers.

To characterize the features of showers induced by different primary particles,
 a parameter named {\it compactness} was invented  by the HAWC
collaboration~\cite{Compactness}.
The {\it compactness} is defined  as $C=N_{hit}/{\rm Max}(Q_i;r>R_c)$, where the maximum function finds the brightest unit outside
an area of radius $R_c$ centred at the shower  core.
The lack of hot spots in the shower, in case of gamma rays,  makes the compactness  $C$  obviously larger
than that of most background cosmic rays with the same $N_{hit}$. For WCDA-1 at the altitude of 4410 m, $R_c$ of 45 meters is found to be optimal, because the electromagnetic component of showers becomes dim enough beyond $R_c$  to make any muon signal outstanding. By cutting on $C>10$, the cosmic ray background can be significantly reduced. This is shown in  Fig.\ref{nhit-distr}, where the red data points show the $N_{hit}$ distribution for all triggers (after filtering out random noise hits as described above) while the black data points are for triggers with $C>10$, rejecting much of the cosmic-ray background. In the range of $N_{hit}<$ 550, about 97.2\% of cosmic ray events are rejected. The rejection power gradually improves  with the event size and reaches 99.8\% for large showers having 700 hits.
To find a suitable threshold for the parameter $C$, the Crab Nebula can be used as a standard candle by maximizing the significance of its detection.

\subsection{The Crab Nebula Detection and WCDA-1 Resolution and Sensitivity}

The Crab Nebula is one of the strongest and most widely studied gamma-ray sources. Its position, angular extension, flux and
Spectral Energy Distribution (SED) have been studied in detail.
The coordinates of the Crab Nebula in the equatorial system (RA, dec) are  $(83.63^\circ, 22.02^\circ)$. The angular extension is much less than 0.07$^\circ$ in the TeV regime.
Events measured by WCDA-1 within an area of 6$^{\circ}$~$\times$~6$^{\circ}$, centered at the Crab nominal position,
have been collected and filled in a grid of $0.1^{\circ}\times 0.1^{\circ}$,
after converting the arrival direction of the shower from local coordinates $(\theta, \phi)$, zenith and azimuth angles, and trigger time to equatorial coordinates. The background has been suppressed as described above. Here, we tuned the threshold to be $C>10$.

In order to calculate the excess of signals from the Crab Nebula direction, the remaining background of the cosmic rays
is estimated using two different methods for cross-checking.
The first one is the  so-called `direct integration' method, developed by the MILAGRO
collaboration~\cite{direct_integration}.
This method is based on the assumption that acceptance of the detector, in local coordinates, is
independent of the trigger rate over a period of several hours and that the cosmic ray background is isotropic.
The last assumption is well justified by the low-energy range considered in this analysis.
The acceptance map is  properly normalized and transformed into an
efficiency map, which is applied to the events that passed the cuts, as measured at time $t$.

An alternative method is the so-called  `equal zenith angle method', developed by the AS$_\gamma$
collaboration~\cite{ASG-EZA}.
This method also relies on the assumption of isotropy of the cosmic ray background. For each
bin $(t,\theta,\phi)$  where the source is supposed to be, the background is integrated over all other bins of the same size at nearby azimuthal angles and the same zenith angle as the source bin.
In this analysis, the background is integrated over 6 azimuthal bins around the one with the source. Therefore, the observation time ratio between the source and background is 1/6.
The underlying assumption for this method is that the detector has a constant detection efficiency for showers
arriving at the same zenith angle.
The bin size is defined by the Point Spread Function
(PSF) of the detector, which is well described by a 2-dimensional Gaussian functional form with the angular resolution defined as  $\Psi_{68}\sim 1.51\sigma=0.84^\circ$, where $\sigma$ is the standard deviation of the Gaussian function and $\Psi_{68}$ is the angle of a cone with 68\% of the probability of the distribution contained, for the lowest energies in this analysis.
The cross check of background estimation between the two methods showed no noticeable difference. Both the event map and background map in the grid have to be smoothed using the PSF.
The excess map is then obtained by subtracting the background map from
the event data map.

\begin{figure}
\centering
\subfigure[]{\includegraphics[width=6cm]{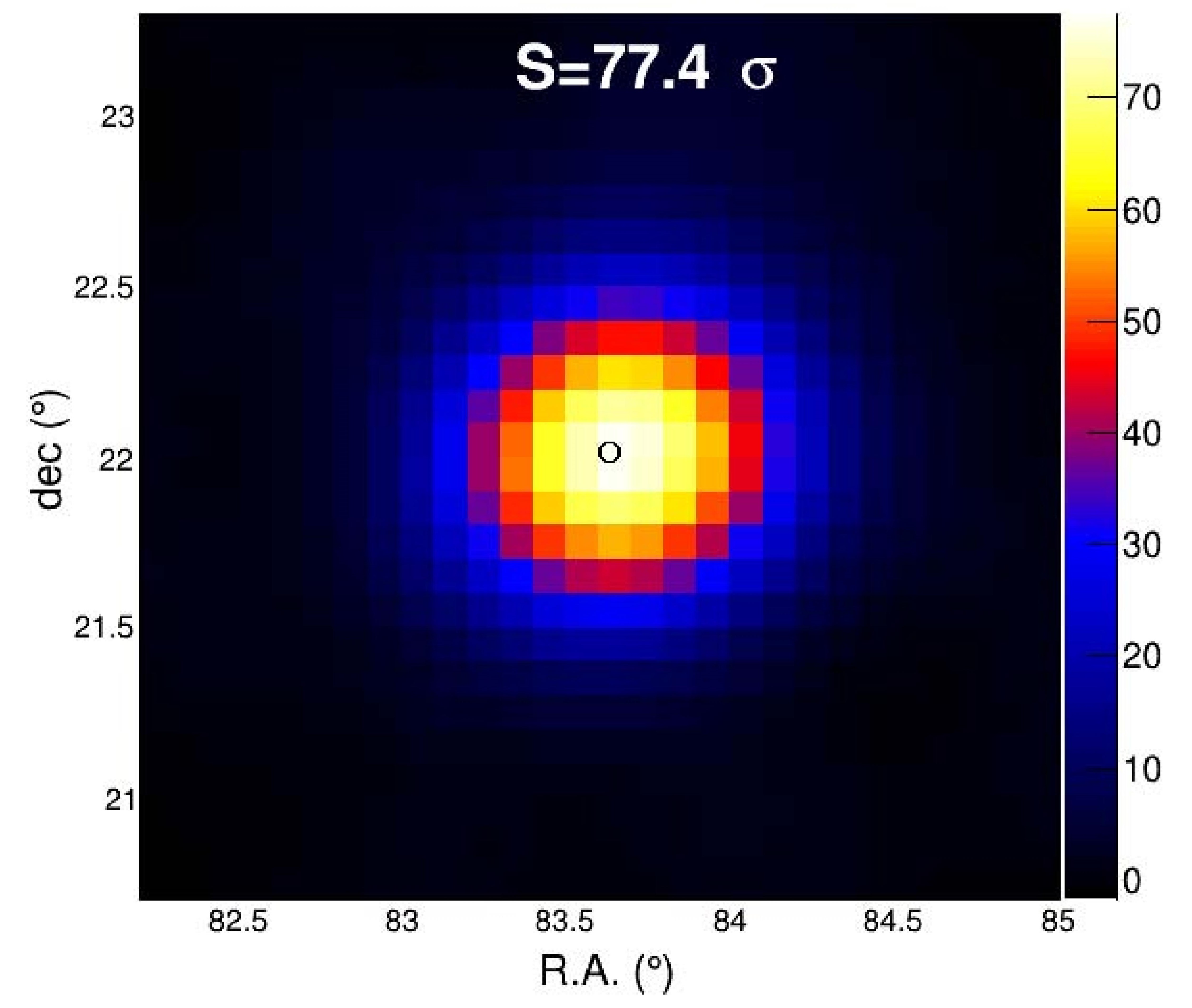}}
\subfigure[]{\includegraphics[width=8cm]{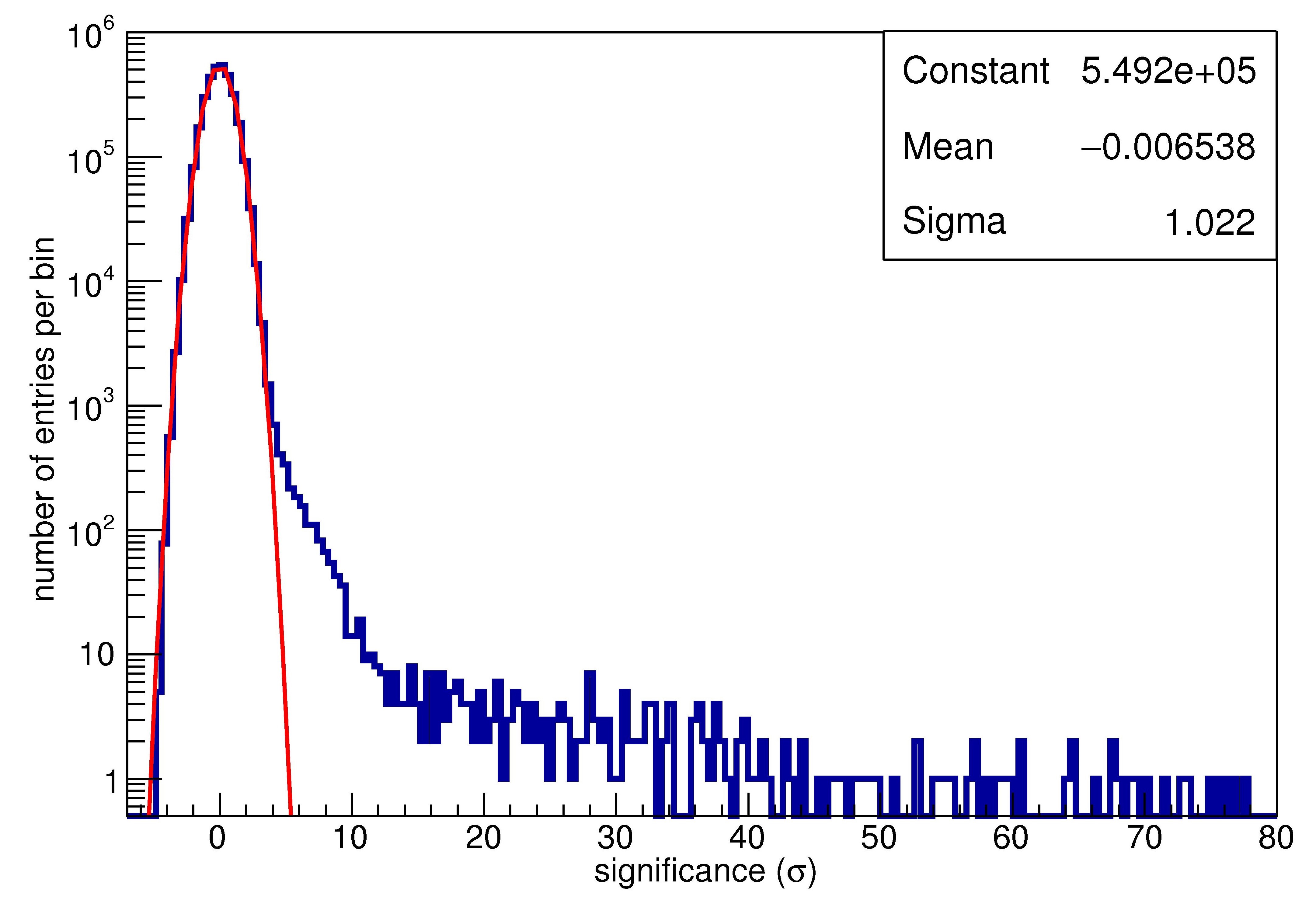}}
\caption{The 2-dimensional significance map around the position of The Crab Nebula (left) and the  distribution of the significance among  $3600\times 900$ bins (right). Measured background cosmic ray events well distributed along a standard normal distribution and the gamma ray signals are clearly distributed over a wide region from $\sim 4 \sigma$ to  77 $\sigma$. }
\label{Sig-map}
\end{figure}

The significance map around the  Crab Nebula and the distribution of the events as a function
of the significance are shown in Fig.~\ref{Sig-map}. The significance is estimated following the method used in\cite{ARGO-survey}.
A clear Crab image with a significance of 77.4$\sigma$ has been achieved. In order to calculate the integrated sensitivity of WCDA-1, all events that have $N_{hit}>100$ collected
up to Feb. 29, 2020 are used including those before Sept. 5, 2019. An integrated sensitivity of 65
milli-Crab-Unit (mCU) per year above 1 TeV (see below) is confirmed by the observation.
The position of the image center at R.A.=$83.65^\circ\pm0.04^\circ$ and Dec=$22.05^\circ\pm0.04^\circ$ indicates that the pointing accuracy of WCDA-1, as a gamma ray telescope, is better than $0.05^\circ$.

\begin{figure}[t]
\centering
\includegraphics[width=0.95\linewidth]{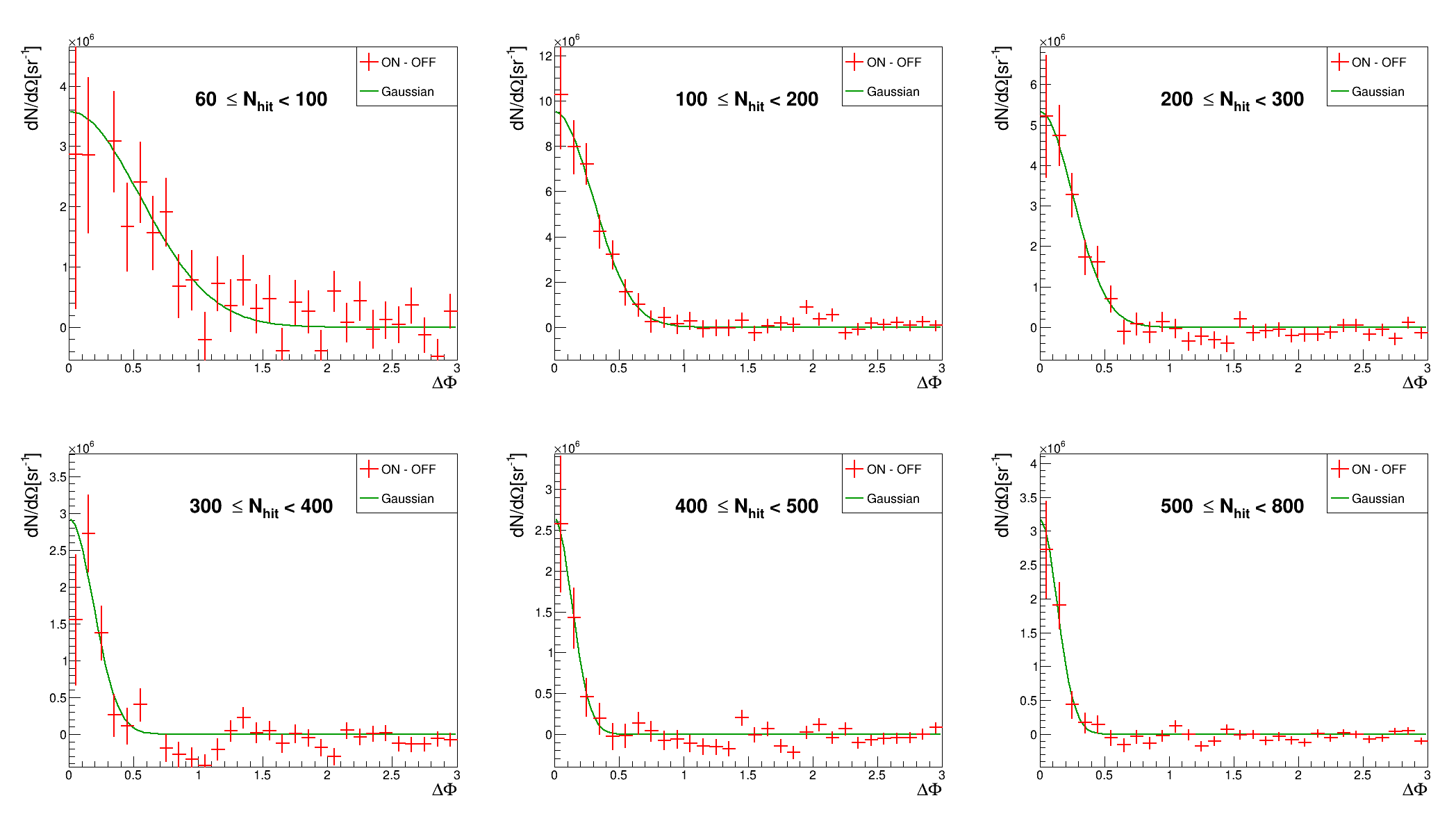}
\caption{Distribution of excess events as a function of the space angle with respect to the Crab Nebula direction in the six bins of
$N_{hit}$. The solid curves indicate  2D-Gaussian fits to the excesses, which result in the angular resolution of 0.84$^\circ$,0.45$^\circ$,0.39$^\circ$,0.29$^\circ$,0.21$^\circ$ and 0.20$^\circ$ in the six bins, respectively.  }
\label{excess}
\end{figure}

The  PSF width of WCDA-1 depends on the shower size, i.e. the shower arrival direction is better measured for larger  showers since the shower front is smoother with larger charge being detected at more sampling points. Therefore, all events collected in the period from Sept. 5, 2019 to Feb. 29, 2020 around the Crab Nebula are grouped into six bins of $N_{hit}$, namely [60,100],[100,200],[200,300],[300,400],[400,500] and [500,800]. Fig.\ref{excess} shows the excesses in the six bins as a function of the space angle between the events and the location of the Crab Nebula. This gives a clear measurement of the PSF in terms of the space angle of the cone in which 68\% of events are contained as a function of shower size, namely 0.84$^\circ$,0.45$^\circ$,0.39$^\circ$,0.29$^\circ$,0.21$^\circ$ and 0.20$^\circ$ in the six bins, respectively. Events with $N_{hit}>800$ give a measure of PSF of 0.20$^\circ$. Events with $N_{hit}<60$ are too small to be reconstructed reliably and are not  taken into account in the analysis reported in this paper.

\section{Monte Carlo Simulation for Gamma Shower Energy Reconstruction  and WCDA-1 Acceptance Estimation}

\subsection{Air Shower Simulation}

The air shower events are generated using  CORSIKA v75000~\cite{corsika}.
Hadronic models EPOS-LHC~\cite{epos} and
FLUKA~\cite{fluka} are selected for the interactions above and below 100 GeV, respectively. All shower particles have been tracked down to
the threshold energy of 50 MeV for hadrons and muons,
and of 0.3 MeV for pions, photons and electrons~\cite{li2014}.
Five primary cosmic ray mass groups, i.e., H, He, C-N-O, Mg-Al-Si
and Fe, are simulated in the background estimation. The Poly-gonato
model~\cite{Horandel2003} is used for the fluxes and spectral indices of the species. The energy range of the incident cosmic rays is assumed to be from 10~GeV to 1~PeV for
protons and helium and from 100~GeV to 1~PeV for the remaining three groups. Cosmic rays are assumed to isotropically arrive in
the zenith angle range from  0$^{\circ}$ to 60$^{\circ}$.
The energy spectrum of gamma rays is sampled from a power law SED with the index of 2.71 for the signals from the Crab Nebula.
Every air shower event generated using CORSIKA is initiated at 100 random locations in an area of  $2000\times2000\rm\;m^{2}$ centered at
the center of WCDA-1 and followed by the detailed simulation of the response of WCDA-1. More than 1.5$\times10^{8}$ comic ray events and
2.7$\times10^{8}$ gamma ray events have been generated in total.

\subsection{Detector Response Simulation for WCDA-1}

\begin{figure}[htb]
\centering\includegraphics[width=6cm]{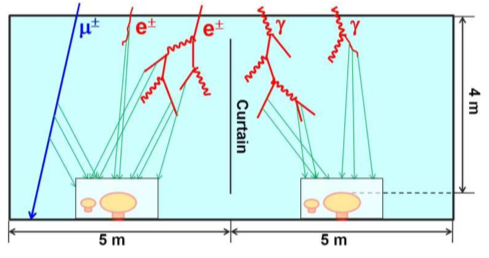}
\centering\includegraphics[width=6cm]{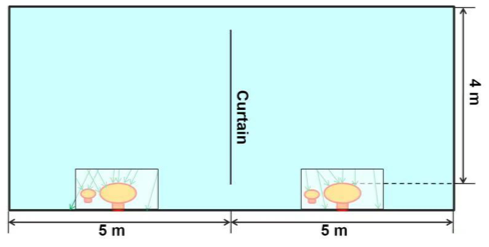}
\caption{Sketch of the two step simulation procedure in G4WCDA. Left panel:
Cherenkov photons/secondary particles were traced in water until entering into the box in
the vicinity of PMTs. Right panel: Only photons or secondary particles that survived the attenuation
cut were further traced until reaching the photocathodes of PMTs.}
\label{g4wcda_2step}
\end{figure}

The code G4WCDA based on GEANT4~\cite{GEANT4} has been developed to simulate the water Cherenkov detector response in the LHAASO experiment.  In order to investigate the
detector performance and correctly estimate the acceptance of WCDA-1, the detailed geometry and structure of the pond, including its roof, steel structure, beams, columns, concrete walls, detector suspension structure and water transparency  are all taken into account in G4WCDA.
Since about 300 Cherenkov photons are produced by an electron or a muon per centimetre of their tracks
in water,  most of the CPU time is consumed to simulate the huge
number of Cherenkov photons for their propagation, including scattering and absorption in water.
The propagation process is completed in two sequential steps as sketched in Fig.~\ref{g4wcda_2step} to improve the simulation efficiency.
In the first step, photons and secondary particles in sub-showers
 are traced
without taking into account the attenuation due to scattering or absorption.
Only for those that entered into a small volume of $80\times 80 \times 80~{cm^{3}}$ around the PMTs in each detector unit,
all parameters of the photons and secondary particles are collected and stored in a ROOT file.  In the second step, the attenuation is taken into account for the stored photons.
The surviving photons and the secondary particles and newly generated photons are further traced until they reach the photocathodes of PMTs.
This approach has proven very effective in coping with various detector operation
conditions, especially those associated with the water transmission which was changing in the beginning phase of operation of WCDA-1.

\begin{figure}
\centering
\subfigure[]{\includegraphics[width=8cm]{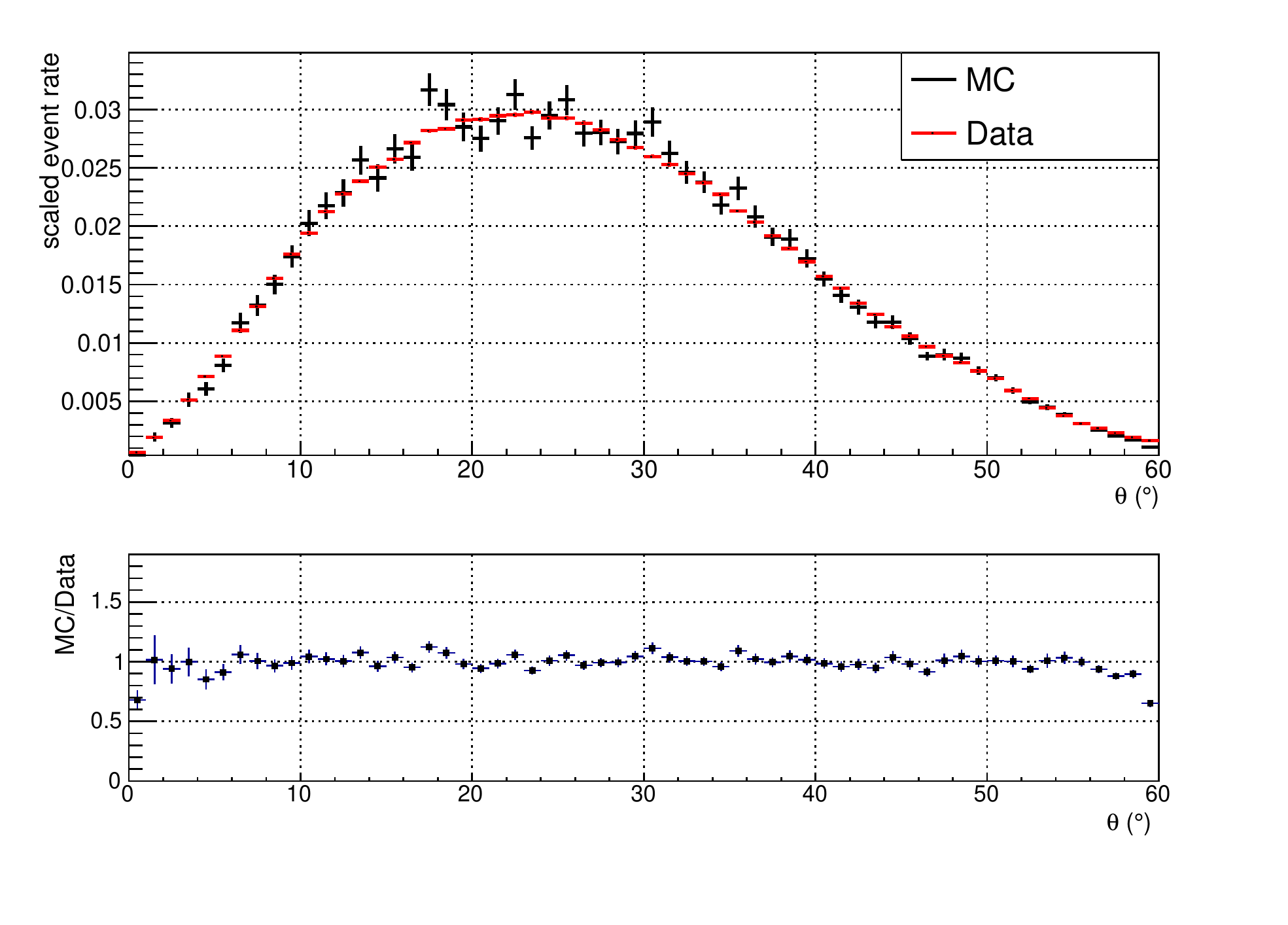}}
\subfigure[]{\includegraphics[width=8cm]{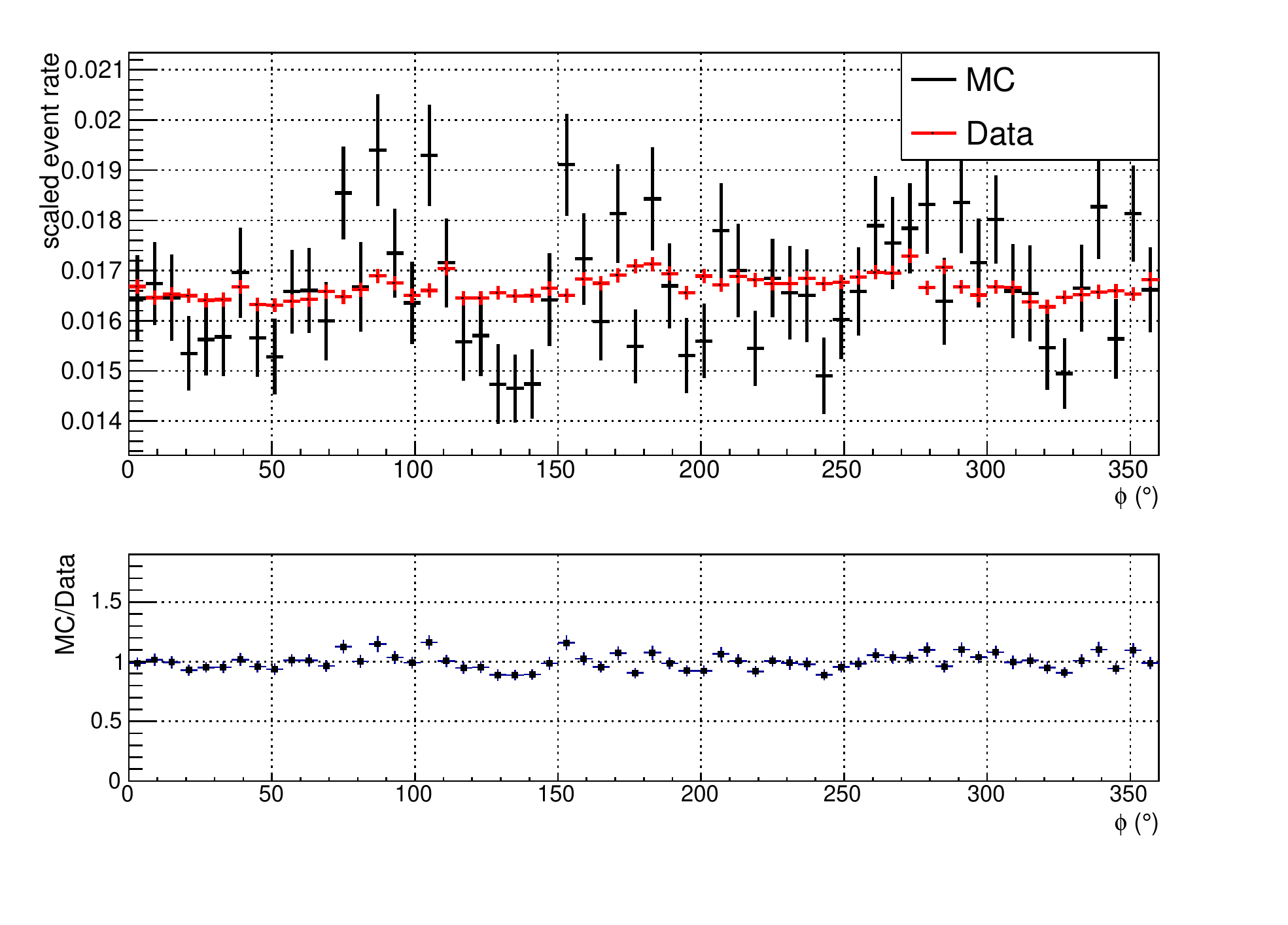}}
\subfigure[]{\includegraphics[width=8cm]{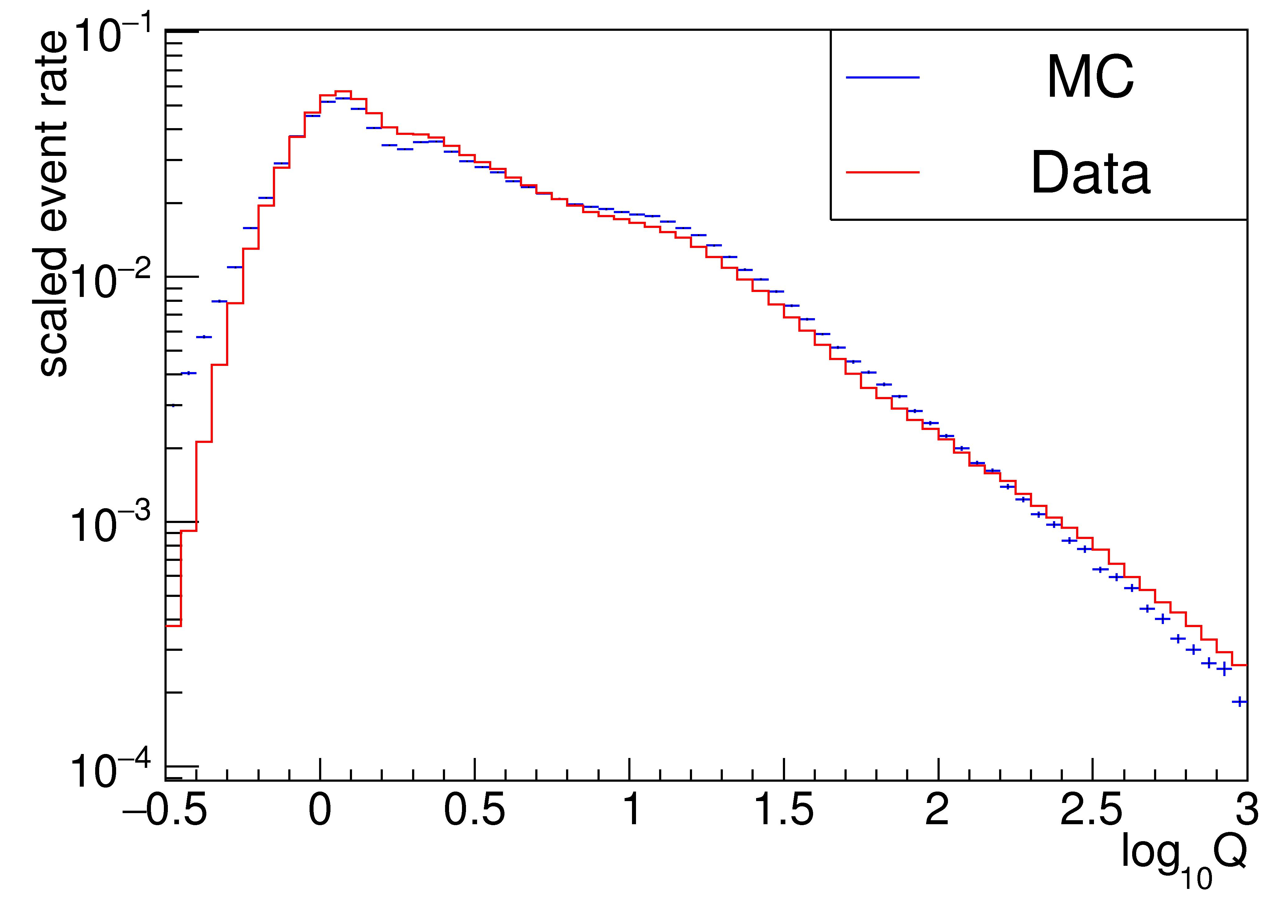}}
\subfigure[]{\includegraphics[width=8.5cm]{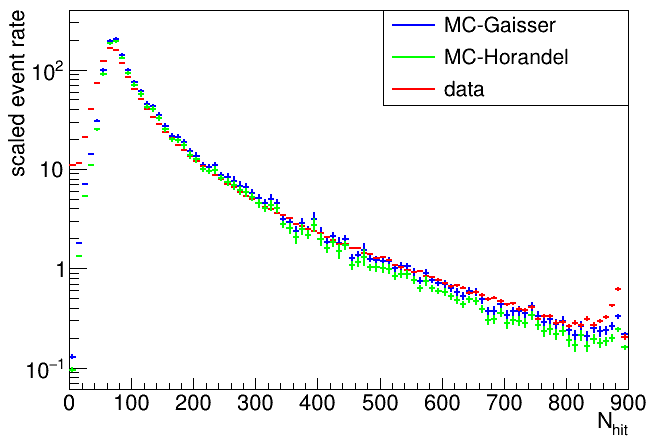}}
\caption{Comparison between simulated (MC) and measured distributions (a) the distribution of
   zenith angle, (b) distribution of azimuth angle,
 (c) distribution of total charge in PE and (d) the distribution of
number of hits $N_{hit}$ in the shower front. The vertical axes are event rates in arbitrary units. In d), simulation results using two composition models due to Horandel\cite{Horandel2003} in green and Gaisser {\it et al.}\cite{gaisser} in blue are compared with the data.}
\label{mc_exp_compare}
\end{figure}

The simulation is tested for its validity by comparing the distributions of
several shower parameters with the measured in the observation.
In Fig.~\ref{mc_exp_compare}(a) and (b), comparisons  of the zenith
($\theta$) and azimuth ($\varphi$) angular distributions between the simulated and measured showers with
 $\theta<50^{\circ}$ are shown. Good agreement is evident,  indicating that the  attenuation of
showers in the atmosphere as well as the uniform response of
the detector array are well described.
Besides the variables related to the arrival direction, two shower energy-related variables
are also compared as  shown in panels (c) and (d) of Fig.\ref{mc_exp_compare}, i.e., the  total charge $Q$ and  $N_{hit}$ recorded by PMTs, respectively.

The shower core reconstruction described in Sec.{\rm\ref{geo-reconstruction}} has been verified using the simulated events. The resolution is found to be 15 m, i.e. a circle with radius of 15 m contains 68\% of the content of the distribution of distances between the thrown and reconstructed core locations, for showers having more than 60 hits in the shower front.

\begin{figure}
\centering
\includegraphics[width=8cm]{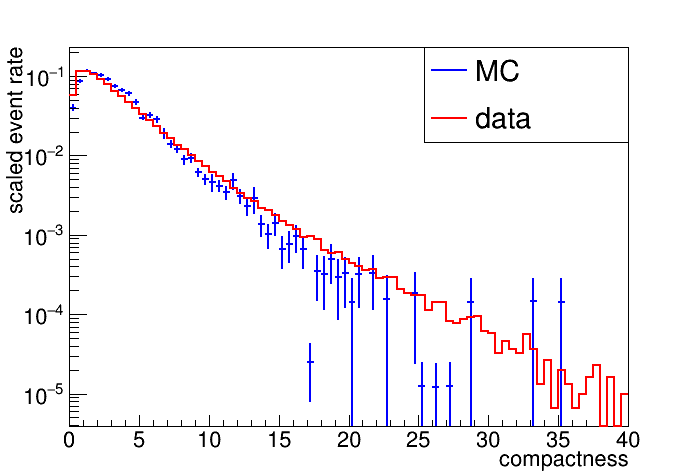}
\includegraphics[width=7.5cm]{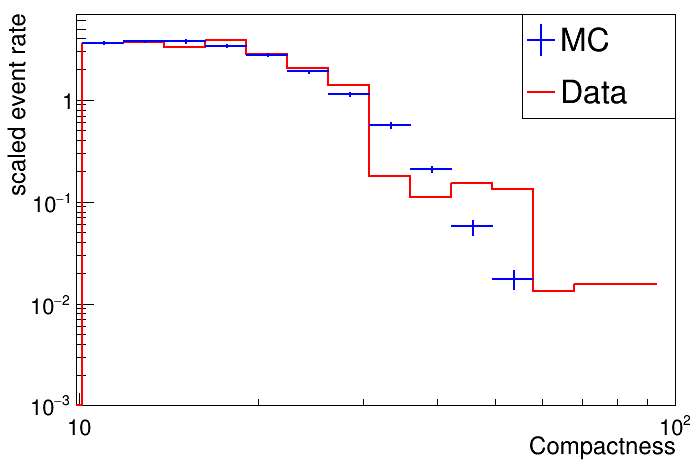}
\caption{Left panel: Distributions of compactness for data (in blue) and  simulated cosmic ray background events (in red). Right panel: The distributions of compactness for selected gamma-like events and simulated events  assuming a simple power-law gamma-ray spectrum  $\propto E^{-2.71}$. }
\label{compactness}
\end{figure}

In order to estimate the acceptance of WCDA-1 for gamma ray showers, the ratio of gamma ray showers surviving the $C$-cut for cosmic ray background suppression has to be estimated correctly. The  measured and simulated $C$ distributions are compared in the left panel of Fig.\ref{compactness}. The agreement between data and the simulation indicates that the algorithm is valid to be used for the gamma-ray-induced showers from the direction of the Crab Nebula. In the right panel, the $C$ distribution of selected events is compared with that for simulated showers with an assumption of a simple power law gamma ray spectrum  of $E^{-2.71}$. With a cut, e.g., $C>10$, a certain number of gamma ray showers are cut while some cosmic ray background events still survive, thus a signal-to-noise ratio yields a significance of the detection for the Crab Nebula. In this way, the threshold of $C$ is tuned to maximize the significance. Simultaneously, the efficiency of gamma ray detection or in other words the effective area is also estimated in each bin of $N_{hit}$, as listed in Table \ref{crabtable}.

\subsection{Energy Reconstruction for Gamma Ray Induced Showers }
\label{sec:E-recons}
\begin{figure}
\centering\includegraphics[width=8cm]{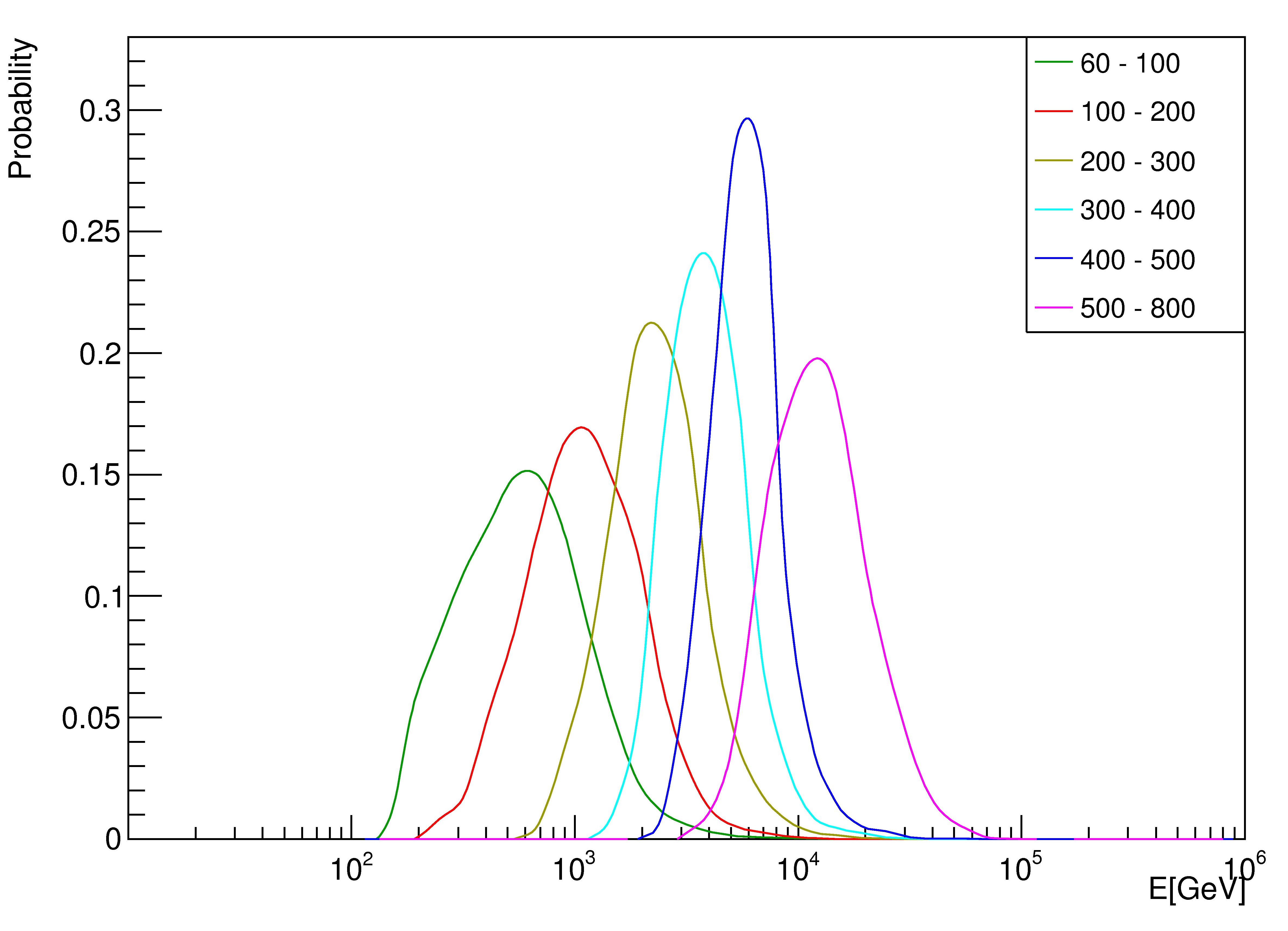}
\caption{Normalized distributions of primary energy of gamma-ray showers
  from a source with spectral index of -2.71.  The color scheme for curves is illustrated in the legend in the figure corresponding to 6 bins of $N_{hit}$.}
\label{mc_energy}
\end{figure}

As mentioned above, the number of hits found within 30 ns of the reconstructed shower front, $N_{hit}$, is selected  as the shower energy estimator. For gamma-induced pure electro-magnetic showers, the simulated showers are used
to establish the correlation between the primary energy and $N_{hit}$.
In Fig.\ref{mc_energy}, primary energy distributions for gamma ray showers in the 6 bins of
$N_{hit}$ are plotted.  Broad and quite overlapping
 distributions are observed.
This indicates that the energy resolution of the detection with WCDA-1 is not perfect, in particular for showers
whose major part may not be contained in WCDA-1. With WCDA-2 and WCDA-3 merged in, the detector sensitive area
 will be large enough to allow more strict event filtering, and the resolution could be improved for the low energy showers in particular. Here, the medians of the distributions are used as the measure of the gamma ray energy for showers in the corresponding bins of  $N_{hit}$ as listed in Table \ref{crabtable}. The energy resolution could be defined well as a symmetric Gaussian function above 6 TeV where the resolution is about 33\%. For lower energies, the resolution of $log_{10}E$ is better defined by a Gaussian distribution, e.g. with a width of 0.5 decades around  1 TeV .  For showers having more than 800
detector units registered, the pile-up effect as shown in Fig. \ref{nhit-distr} makes the differential flux
measurement impossible due to the natural saturation of $N_{hit}$ at 900. An integrated flux above the energy
corresponding to $N_{hit}=800$ could be in principle reported, however, the statistical significance is too weak, $<2\sigma$,
to be reliable for the flux measurement here. Moreover, including a large number of units with too much charge recorded would
require more careful tuning in the simulation to better simulate the response of the detector, as indicated in
Fig. \ref{mc_exp_compare} (d). The high energy events include both gamma-ray and cosmic ray induced showers,
thus requiring a better energy estimator than $N_{hit}$.

\begin{table}[htb]
\centering
\caption{Summary of data used in the measurement of SED of the Crab Nebula over $3.57\times10^{6}$ seconds. }
\label{crabtable}
\small
\begin{tabular}{llccccc}
\hline
\hline
& N$_{hit}$ &$E_{med}$   & Excess & Background&Significance& Differential Flux \\
&           &  $ ({\rm TeV})$ &       &       &    ($\sigma$) & $(cm^{-2}s^{-1}TeV^{-1})$\\
\hline
(a)& ~60 - 100  & 0.58 &  1438.2 & 24885.8  & 9.1& $(1.66\pm 0.20 ) \times 10^{-11}$  \\
(b)& 100 - 200  & 1.1  &  1082.7 &  5202.3  &15.0& $(2.89\pm 0.23 ) \times 10^{-11}$ \\
(c)& 200 - 300  & 2.4 &    456.2  & 1376.8  &12.3& $(4.74\pm 0.48 ) \times 10^{-12}$ \\
(d)& 300 - 400  & 3.9 &    161.2  &  335.8  &8.8& $(1.12\pm 0.17 ) \times 10^{-13}$ \\
(e)& 400 - 500  & 5.9 &     60.3   &  77.7  &6.8& $(3.54\pm 0.74 ) \times 10^{-13}$ \\
(f)& 500 - 800  & 12.1 &    82.7   &  45.3 &12.3&$(6.91\pm 1.0 ) \times 10^{-14}$ \\
\hline
\end{tabular}
\end{table}

\section{Spectral Energy Distribution of the Crab Nebula}\label{Sec:SEDCrab}

\begin{figure}
\centering
\subfigure[]{\includegraphics[width=9cm]{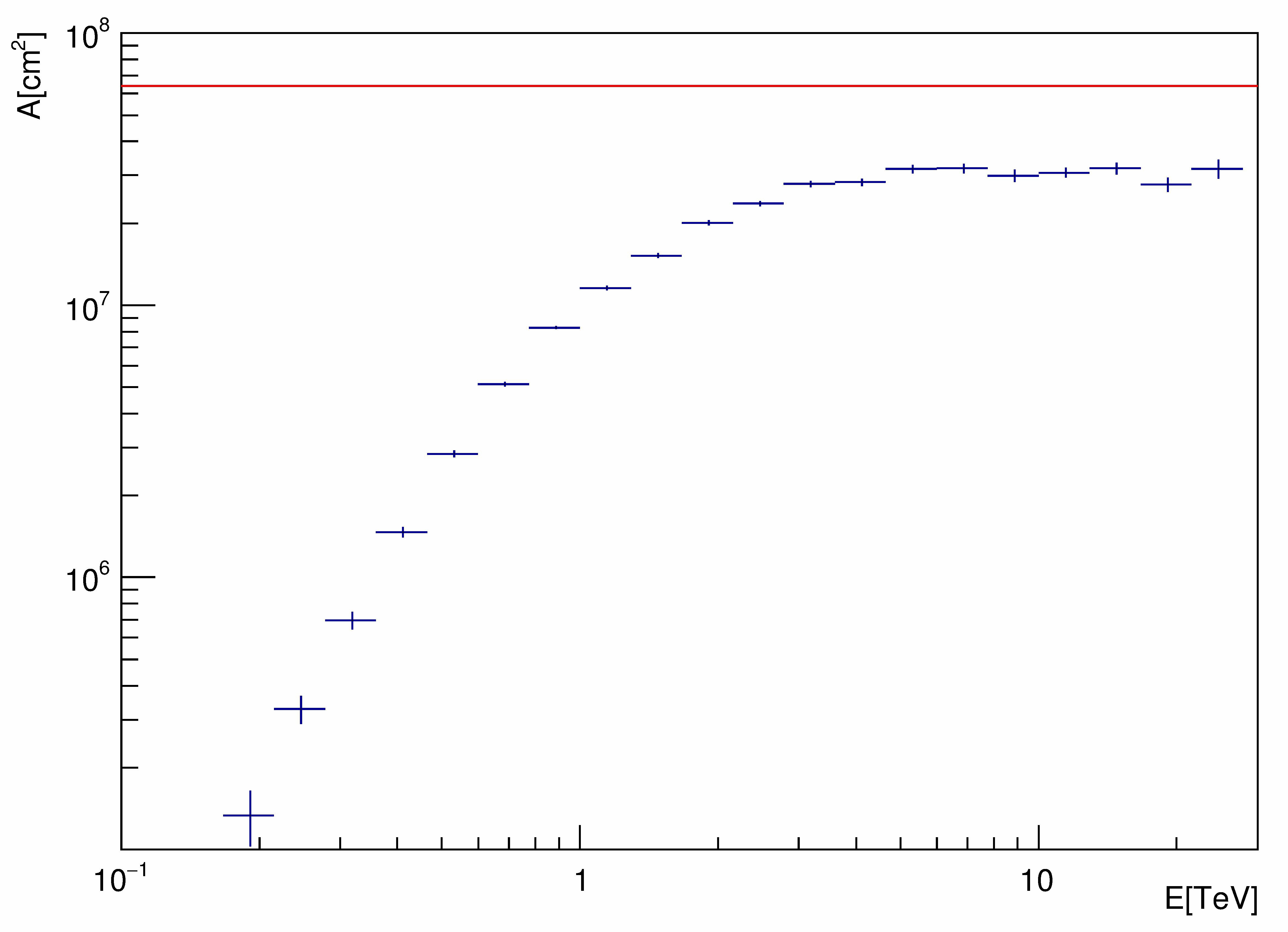}}
\caption{The effective area of WCDA-1 for the Crab detection with the cuts applied in the analysis according to the simulation for gamma rays. The red line indicates the physical area of the WCDA-1 used in the analysis.}
\label{eff-area}
\end{figure}

The SED of the Crab Nebula  is measured using data collected from  Sep. 5, 2019 to Feb. 29, 2020,
for a live time of $3.57\times10^{6}$ seconds.
The criteria for events to be used in  the analysis are as follows.
\begin{itemize}
\item the core is within a square of 80m$\times$80m centered at the WCDA-1 center.
\item $60 \le N_{hit}<800$.
\item The Crab Nebula was at least 45$^\circ$ above the horizon.
\item The compactness is larger than 10.
\end{itemize}
All the criteria helped to maximize the detection significance. The corresponding significance, number of
events exceeding the background and remaining cosmic rays in each bin are listed in  Table~\ref{crabtable}. Under the cuts, WCDA-1 has an effective area that varies from about 400 $m^2$ at 600 GeV to a constant of 3200 $m^2$ above 4.5 TeV as plotted in Fig. \ref{eff-area}. Importantly, this indicates that above 4.5 TeV the SED measurement has a constant efficiency of about 50\% mainly due to the gamma selection requiring $C>10$.

The SED is determined by minimizing the $\chi^2$ function
\begin{eqnarray}
  \chi^{2}= \sum_{i=1}^6\frac{\left[N_{i}^{obs} - N_{i}^{exp}(\phi_{0},\alpha,\beta)\right]^2}{(\sigma_{i}^{obs})^2}
\end{eqnarray}
\noindent where $N_{i}^{obs}$ is the number of events exceeding the background  in the $i$-th
bin of $N_{hit}$,  $\sigma_{i}^{obs}$ is the statistical error of $N_{i}^{obs}$  and
$N_{i}^{exp}(\phi_{0},\alpha, \beta)$ is the  expected number of events according to the hypothesis of a log-parabolic model
\begin{eqnarray}
\phi(E)=\phi_{0}\left(\frac{E}{\rm 3\ TeV}\right)^{-\alpha-\beta~\rm ln \left(\frac{E}{\rm 3\ TeV}\right)}
\end{eqnarray}
with three parameters ($\phi_{0}$, $\alpha$, $\beta$), where $E$ is shower energy and $\phi$ is the photon flux from the Crab Nebula.
This formula was suggested as one of the best functional approximations for the
Crab nebula spectrum over a wide energy range\cite{log-prob}.
The number of events for signals  from the Crab Nebula direction is shown in Fig.\ref{excess} for each bin of $N_{hit}$ with the background cosmic ray events subtracted. They can be fit quite well with a 2-dimensional Gaussian functional template.

\begin{figure}
\centering
\includegraphics[width=9cm]{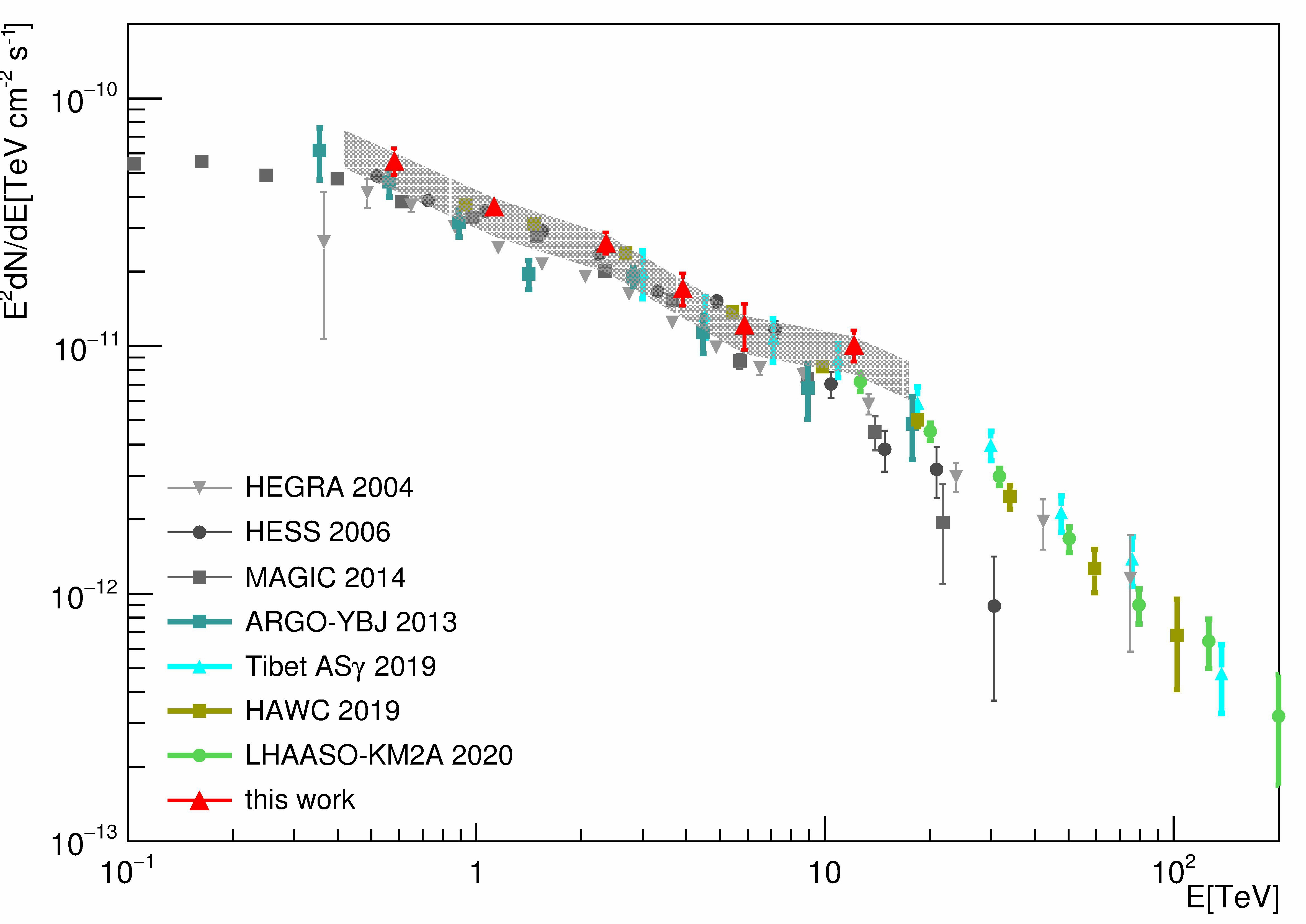}
\caption{The spectral energy distribution of the Crab Nebula in the energy range from 800 GeV to 13 TeV
   measured by WCDA-1 of LHAASO is shown as the red filled triangles. As a comparison, the SED measured by using KM2A of LHAASO\cite{crab-KM2A} above 10 TeV is also shown as the green filled circles. The  error bars are statistic errors and shade area represents the systematic error. Measurements of other experiments\cite{ARGO-Crab}\cite{ASG-Crab}\cite{HAWC-Crab}\cite{HESS-Crab}\cite{MAGIC-Crab}
are also plotted for comparison.}
\label{crabspectrum}
\end{figure}

The SED measured using WCDA-1 is shown in Fig.~\ref{crabspectrum} as the red points. The corresponding  differential flux in each bin is listed also
in Table \ref{crabtable}.
he spectral parameters obtained in the log-parabolic fitting are  $\phi_{0}=(2.32\pm0.19)\times10^{-12}cm^{-2}~TeV^{-1}~s^{-1}$,
$\alpha=2.57\pm0.06$,  and $\beta=0.02\pm0.05$ with $\chi^{2}/ndf$=0.513, respectively.

For comparison, the SED measured by other ground-based gamma ray experiments,
i.e., HAWC~\cite{HAWC-Crab}, Tibet AS$_\gamma$~\cite{ASG-Crab} and ARGO-YBJ~\cite{ARGO-Crab}, are
shown in the same figure, together with the measurements by the IACT experiments, HESS~\cite{HESS-Crab} and
~MAGIC\cite{MAGIC-Crab}.

The  systematic errors are mainly caused by following issues.
Firstly, WCDA-1 has been continuously improving in terms of the water transparency and detector stability during
the observation reported in this paper. This is reflected by Fig.\ref{muon-calib} (b) that shows the measured
average PMT charge generated by single muons nearly monotonically increased with time except during major
maintenance in October 2019. This requires a correction of about $\pm$11\% to the charge measurements, thus
introducing an uncertainty in the detection efficiency for events with a small number of hits,
namely about 8.8\% at $Q=0.3$~PE, including non-uniformity among 900 detector units and variation over 170
operational days, tested by using cosmic rays which are uniformly and  hitting  the detector from all directions..
Secondly, the shower and detector simulation is an essential tool in estimation of the detector acceptance
in terms of effective area for gamma ray detection. There is no way to check its validity directly using
gamma rays since we do not have a known gamma ray beam for the testing. However, it can be checked  using
cosmic rays. Assuming that the central region of WCDA-1, e.g., a circle with radius of 30 m, is fully efficient
for large enough showers, e.g., showers with more than 500 hits, one can calibrate the absolute flux of cosmic
rays calculated using the simulation codes in comparison with observations. Here, the details about the
cosmic ray composition and the models of interactions with the atmospheric nuclei in the relevant energy
range below 100 TeV have to be assumed as well. As shown in Fig.\ref{mc_exp_compare}, there are discrepancies
between simulated and measured distributions. Here two composition assumptions are compared with data in
the $N_{hit}$ distributions. This results in an uncertainty of the estimate of detector efficiency associated
with the assumptions, including  the entangled issues of interaction modeling and cosmic ray composition.
The larger discrepancy of 16\%  between the two assumptions is taken  as a conservative estimate of the
corresponding uncertainty of the detector efficiency. The uncertainty of gamma ray detection acceptance
estimated by using the simulation tool is assumed to be the same in this analysis.
Thirdly, the gamma ray shower energy scale also totally relies on the simulation. As shown in
Fig.\ref{mc_exp_compare} (c), the simulation of the detector response near the threshold is slightly overestimated.
In order to estimate the associated uncertainty of the energy scale, energies of the gamma events from the
Crab Nebula direction have been reconstructed with different thresholds of $Q_{th}$, 0.3 PE and 0.6 PE,
respectively. A difference of 3\% is found between the  average energies due to the two reconstructions;
thus a uncertainty of 8\% is expected in the SED.
The  overall systematic uncertainty could be as large as $^{+8}_{-24}$\%. It is presented in Fig.\ref{crabspectrum} as the shaded area around the SED measured  by WCDA-1.

\section{Conclusions}\label{Sec:conclusions}

In this paper, the performance of the first LHAASO water Cherenkov detector,
WCDA-1, is tested by observing the Crab Nebula as a standard candle of TeV gamma ray astronomy.
The Crab Nebula has been detected with
a significance of 77.4 $\sigma$, corresponding to a sensitivity of 65 mCU, in agreement with the
design specification.
The angular resolution for the gamma ray arrival direction is found to be better than 0.4$^\circ$ above 3
TeV and the pointing accuracy to be better than 0.05$^\circ$. This is a result of well-measured detector orientation and timing calibration of every detector unit. Air shower and detector simulations are well tuned to reproduce the measured cosmic ray data and used for the gamma-ray shower energy reconstruction and detector acceptance estimation.
The spectral energy distribution of the Crab Nebula has been measured in the TeV range from 0.5 TeV
to 15.8 TeV showing a good agreement with all other major gamma ray astronomy experiments.

\vspace*{1cm}

\hspace*{-6mm} \acknowledgments{{\bf Acknowledgments} The authors would like to thank all staff members who work at the LHAASO site above 4400 meter above the sea level year round to maintain the detector and keep the water recycling system, electricity power supply and other components of the experiment operating smoothly. We are grateful to Chengdu Management Committee of Tianfu New Area for the constant financial support for research with LHAASO data. This research work is also supported by the following grants: The National Key R\&D program of China under grants 2018YFA0404201, 2018YFA0404202 and  2018YFA0404203, by the National Natural Science Foundation of China (NSFC grants No.12022502, No.11905227, No.U1931112, No.11635011, No.11761141001, No.Y811A35, No.11675187, No.U1831208), and in Thailand by RTA6280002 from Thailand Science Research and Innovation.}

\begin{multicols}{2}

\end{multicols}

\clearpage
\end{document}